\documentclass[12pt, a4paper]{article}

\usepackage[top = 2.0cm, left = 2.0cm, right = 2.0cm, bottom = 2.0cm]{geometry}
\usepackage{amsfonts}
\usepackage{amsmath}
\usepackage[english]{babel}
\usepackage{bm}
\usepackage{caption}
\usepackage[T1]{fontenc}
\usepackage{indentfirst}
\usepackage[utf8]{inputenc}
\usepackage{placeins}
\usepackage{setspace}
\usepackage{xcolor}
\usepackage{graphicx}
\usepackage{multirow}
\usepackage{booktabs}
\usepackage{makecell}
\usepackage[superscript,biblabel]{cite}
\usepackage{hyperref}
\usepackage{graphicx}
\usepackage{subcaption}
\usepackage{float}
\usepackage{placeins}
\usepackage{adjustbox}
\usepackage{ltablex}

\newlength{\rowh}
\title{Continuous monitoring of delayed outcomes in basket trials}

\author{
M\'arcio A. Diniz\thanks{Department of Population Health Science and Policy, Icahn School of Medicine at Mount Sinai, New York, NY, USA.}
\thanks{Institute for Healthcare Delivery Science, Icahn School of Medicine at Mount Sinai, New York, NY, USA.}
\thanks{Tisch Cancer Institute, Icahn School of Medicine at Mount Sinai, New York, NY, USA.}
\and Hulya Kocyigit\footnotemark[1] \footnotemark[2] \footnotemark[3]
\and Erin Moshier\footnotemark[1] \footnotemark[2] \footnotemark[3]
\and Madhu Mazumdar\footnotemark[1] \footnotemark[2] \footnotemark[3]
\and Deukwoo Kwon\thanks{Department of Biostatistics, College of Medicine, University of Arkansas for Medical Sciences, Little Rock, AR, USA.}
}

\date{\today}

\begin{document}

\maketitle

\thispagestyle{empty}

\begin{abstract}
Precision medicine has led to a paradigm shift allowing the development of targeted drugs that are agnostic to the tumor location. In this context, basket trials aim to identify which tumor types - or baskets - would benefit from the targeted therapy among patients with the same molecular marker or mutation. We propose the implementation of continuous monitoring for basket trials to increase the likelihood of early identification of non-promising baskets. Although the current Bayesian trial designs available in the literature can incorporate more than one interim analysis, most of them have high computational cost, and none of them handle delayed outcomes that are expected for targeted treatments such as immunotherapies. We leverage the Bayesian empirical approach proposed by Fujiwara et al., which has low computational cost. We also extend ideas of Cai et al to address the practical challenge of performing interim analysis with delayed outcomes using multiple imputation. Operating characteristics of four different strategies to handle delayed outcomes in basket trials are compared in an extensive simulation study with the benchmark strategy where trial accrual is put on hold until complete data is observed to make a decision. The optimal handling of missing data at interim analyses is trial-dependent. With slow accrual, missingness is minimal even with continuous monitoring, favoring simpler approaches over computationally intensive methods. Although individual sample-size savings are small, multiple imputation becomes more appealing when sample size savings scale with the number of baskets and agents tested.

\vspace{0.5cm}

\noindent \textbf{keywords:}basket trial, interim analysis, missing data, multiple imputation
\end{abstract}


\section{Introduction}
\label{sec:intro}

In oncology, novel drugs have historically been developed on the basis of the anatomical site of the primary tumor. Precision medicine has led to a paradigm shift that allows the development of targeted drugs that are agnostic to the location of the tumor. In this context, basket trials aim to identify which tumor types - or baskets - would benefit from the targeted therapy among patients with the same molecular marker or mutation. As a nonrandomized discovery trial design typically used in phase 1b or single-arm phase 2 trials, efficacy in basket trials is often assessed based on a dichotomous endpoint, such as the objective response rate (ORR) at 3 months, compared to a historical control. To date, the U.S. Food and Drug Administration (FDA) has approved six tumor-agnostic therapies, with two development programs based on a basket trial design. Moreover, Kasim et al. \cite{kasim_basket_2023} identified 138 basket trials registered in ClinicalTrials.gov between 2001 to 2019.  

As the sample size for each basket is small, frequentist and Bayesian designs that borrow information across baskets have been developed. Cunanan et al. \cite{cunanan_efficient_2017} proposed a frequentist pruning-and-pool approach that performs an interim analysis to drop baskets due to futility in the first stage, then pooling the promising ones in the final analysis. Thall et al. \cite{thall_hierarchical_2003} introduced a Bayesian hierarchical modeling approach assuming that response rates of different baskets are derived from the same distribution, known as exchangeability assumption, which has been further relaxed to accommodate high heterogeneity between baskets \cite{neuenschwander_robust_2016, chu_blast_2018}. Psioda and Xu \cite{psioda2021bayesian} presented a Bayesian model averaging approach considering all possible models for the response rate ranging from all baskets having the same probability of response to each basket having a different probability of response, generalizing the ideas initially introduced by Simon et al. \cite{simon_bayesian_2016}. More recently, Fujiwara et al. \cite{fujikawa_bayesian_2020} proposed an Empirical Bayesian approach that combines posterior parameters based on the similarity between posterior distributions measured by Jensen-Shannon Divergence.

 In the context of standard single-arm phase 2 trials, continuous monitoring has been proposed to increase the likelihood of early termination for futility \cite{thall1994practical}. However, one of the main challenges to conducting continuous monitoring is that the response assessment might not be readily available for all currently enrolled patients in each interim analysis. Suspending trial accrual until complete data are obtained for all patients would be the recommended strategy, although it is not practically feasible because it would significantly increase the duration of the trial. Consequently, developing efficient statistical methods that can effectively handle these delayed outcomes without compromising the timeliness of critical decision-making remains an ongoing necessity in clinical trial design. Cheung and Tall \cite{cheung2002monitoring} incorporated partial patient information in likelihood by assigning weights proportional to the follow-up period that has elapsed without observing a response, and Cai et al. \cite{cai_bayesian_2014} proposed using multiple imputation based on a piecewise exponential survival model to incorporate the uncertainty of patients with incomplete data.

We propose the implementation of continuous monitoring for basket trials to increase the likelihood of early identification of non-promising baskets. Although the Bayesian trial designs aforementioned can incorporate more than one interim analysis, most of them have high computational cost and none of them handle delayed outcomes that are expected for targeted treatments such as immunotherapies. Thus, we leverage the approach of Fujiwara et al. that has low computational cost, while extending the ideas of Cai et al to address the practical challenge of performing interim analysis with delayed outcomes using multiple imputation based on a Weibull survival model.

The remainder of this article is organized as follows: in section \ref{sec:methods}, we introduce the basket trial design and discuss possible strategies for handling incomplete data during an interim analysis, including steps for multiple imputation; in section \ref{sec:simu}, we outline the simulation setup defining the scenarios; in section \ref{sec:results}, we examine the results of the extensive simulation studies; concluding remarks are discussed in section \ref{sec:discussion}.

\section{Methods}
\label{sec:methods}

We consider $B \geq 2$ baskets, with each basket enrolling $n_b$ patients with a binary response variable modeled as $Y_{bi} \sim \mbox{Bernoulli($\theta_b$)}$ where $\theta_b$ is the probability of response at time window $T$. In each basket, we would like to test the hypothesis that the response rate within period $T$ is greater than $\phi$, in other words, $H: \theta_b > \phi$.

At every $n_b^*$ enrolled patients in the basket $b$, an interim analysis is performed considering conjugate prior distributions $\theta_b \sim \mbox{Beta($s_1, s_2$)}$. The posterior distribution for basket $b$ can be calculated as follow:
\begin{eqnarray}
\theta_b \sim \mbox{($s_1 + r_b, s_2 + n_b - r_b$)}
\label{eq:posterior_nosharing}
\end{eqnarray}
where $r_b = \sum_{i = 1}^{n_b} y_i$ for $b = 1, \ldots, B$ without sharing information between baskets with the underlying assumption that the response variable was observed for all $n_b^*$ patients in a given basket.   

\begin{figure}[!h]
    \centering
    \includegraphics[width=0.5\linewidth]{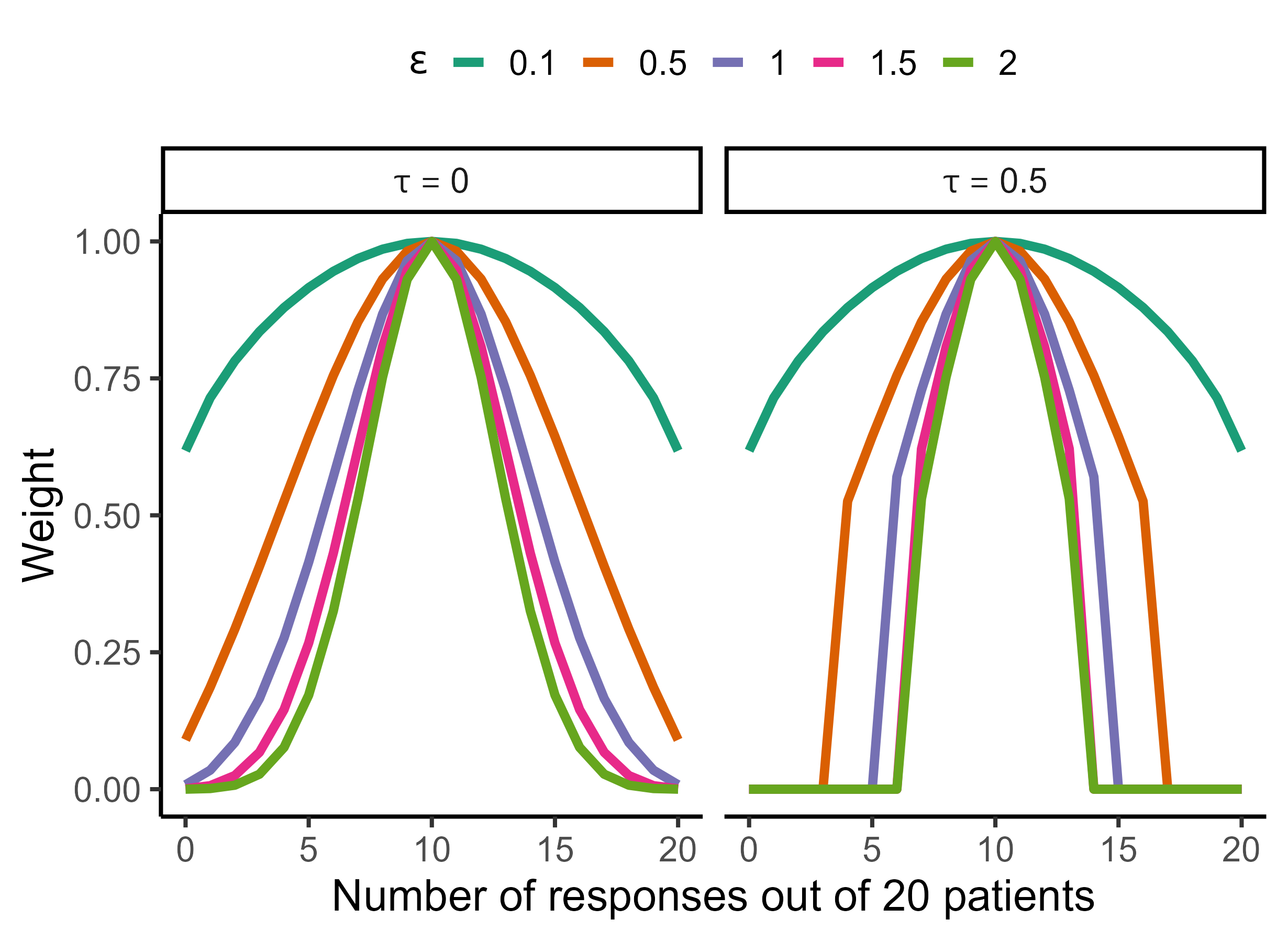}
    \caption{Weights indicating the similarity between two baskets based on Jensen-Shannon Divergence  when 10 responses out of 20 patients is observed in given basket as function of the number of responses out of 20 patients in a basket.  The parameters $\epsilon$ and $\tau$ are tuning parameters.}
    \label{fig:weights_js}
\end{figure}

Following the Empirical Bayes approach introduced by Fujiwara et al \cite{fujikawa_bayesian_2020}, we calculate weights that control the amount of information borrowing across all baskets. These weights are determined by the similarity between the posterior distributions of each individual basket, calculated under the assumption that no information has been shared between them, 

\begin{eqnarray}
w_{bk} = \begin{cases} \left(1 - JSD(\pi(\theta_b|r_b), \pi(\theta_k|r_k)) \right)^\epsilon \quad \mbox{if $\left(1 - JSD(\pi(\theta_b|r_b), \pi(\theta_k|r_k)) \right)^\epsilon > \tau$} \\
0 \quad \mbox{otherwise}
\end{cases}
\label{eq:weights_js}
\end{eqnarray}
where JSD(Q, P) is the Jensen-Shannon Divergence between the distributions P and Q, $\epsilon$ and $\tau$ are a power and a threshold tuning parameters, respectively. Figure \ref{fig:weights_js} shows the weight when 10 responses out of 20 patients are observed in a given basket as function of the number of responses out of 20 patients in another basket, $\tau$ and $\epsilon$ . 

For simplicity, we assume the values of  $\epsilon = 2$ and $\tau = 0$ which result in weights ranging from 0.094 to 1 based on recommendations discussed in Fujiwara et al\cite{fujikawa_bayesian_2020}. Therefore, the posterior distribution for each basket borrowing information across baskets depending on the similarity between them is derived as follows
\begin{eqnarray}
\theta_b \sim \mbox{Beta$\left(\sum_{k = 1}^B w_{bk} (s_1 + r_k), \sum_{k = 1}^B w_{bk} (s_2 + n_k - r_k)\right)$}
\label{eq:posterior_sharing}
\end{eqnarray}

Then, the intervention will be considered futile for basket $b$ if $P(\theta_b < \phi|r_b) > \gamma_{interim}$, where $\gamma_{interim}$ is a design parameter defined based on desirable operating characteristics obtained in simulation studies.

\begin{figure}
    \centering
    \includegraphics[width=0.5\linewidth]{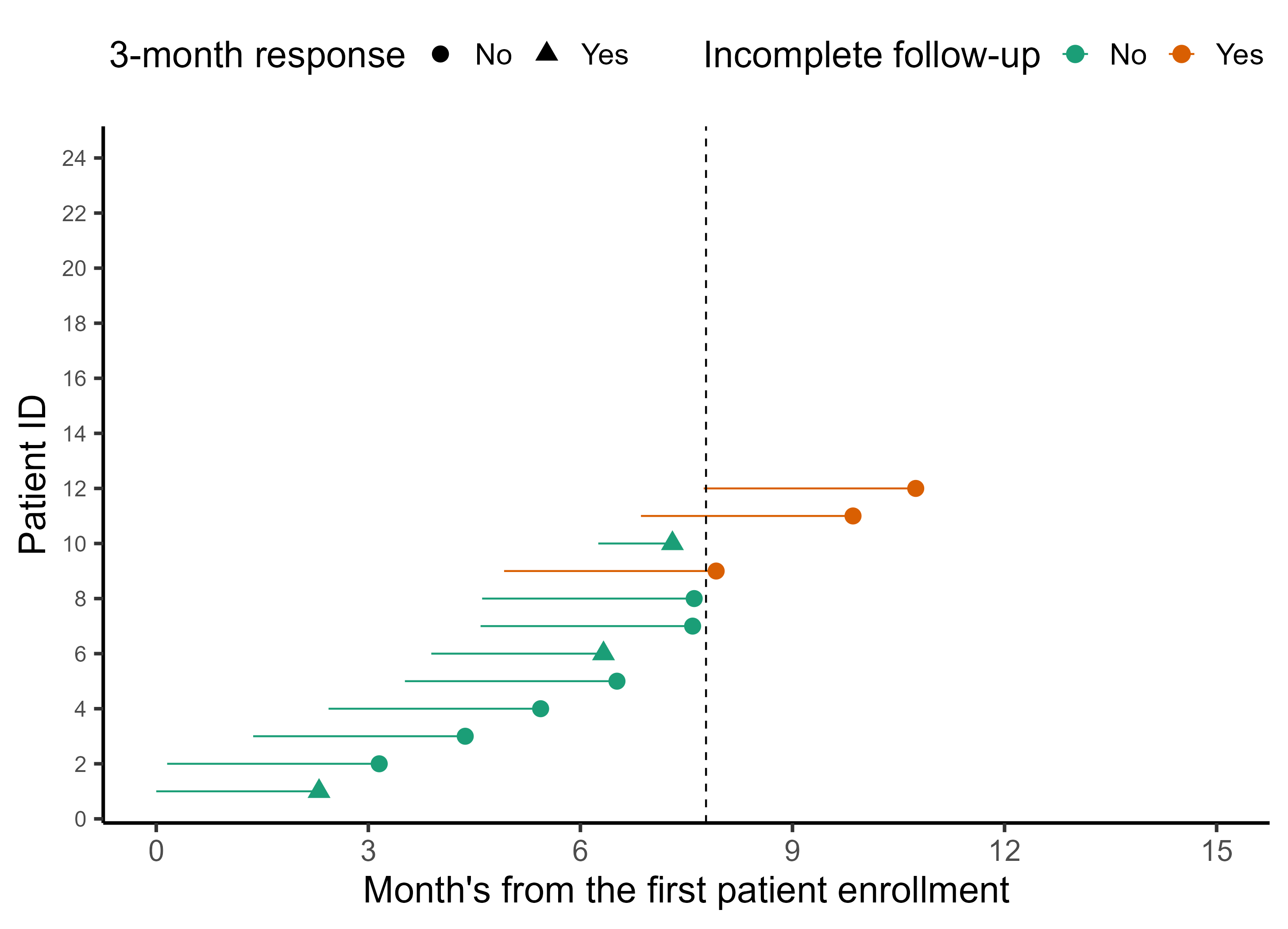}
    \caption{Follow-up status when an interim analysis is performed after 20 patients are enrolled assuming an accrual of 4 patients per month with response variable observed at 3 months. Only 8 patients out 20 have complete 3-month follow-up.}
    \label{fig:interim_analysis_status}
\end{figure}

However, the assumption that the response variable is quickly observed for all enrolled patients up to the interim analysis may be questionable in practice, specially when the time window $T$ for response is long, as often it is considered in trials for immunotherapies. Figure \ref{fig:interim_analysis_status} illustrates the incomplete follow-up of patients at the time of an interim analysis, which is performed after 20 patients have been enrolled. This scenario assumes a patient accrual of four patients per month and a primary response assessment time of three months. 

In such scenarios, the posterior distribution \eqref{eq:posterior_sharing} cannot be calculated without additional considerations.  We leveraged the ideas introduced by Cai et al\cite{cai_bayesian_2014} that has proposed the use of multiple imputation using a Piecewise Exponential distribution with ad-hoc time points, which makes the practical implementation of their approach not very straightforward. Instead, we assumed that the time to response can be modeled using a Weibull distribution, $T_{bi} \sim \mbox{Weibull($\kappa_b, \sigma_b$)}$ where $\kappa_b$ and $\sigma_b$ are the shape and scale parameters, respectively. The observed times are $x_{bi} = \min\{c_{bi}, t_{bi}\}$, where $c_{bi}$ is the actual follow-up time with the response indicator $\delta_{bi}$ for patient $i$ for $i = 1, \ldots, n_b$ and $b = 1, \ldots, B$.
The scale parameter $(\sigma)$ can be re-written within a regression framework according to different modeling strategies and weakly informative priors can be chosen for all parameters. 


When baskets are not considered in the multiple imputation, time to response can be modeled as follows,
\begin{eqnarray}
log(\sigma_b) &=& \beta_0  \nonumber \\
\beta_0 &\sim& \mbox{Normal($0, 100$)}  \nonumber \\
\kappa &\sim& \mbox{Uniform($0, 10$)}.
\end{eqnarray}
where $b = 1, \ldots, B$ and $i = 1, \ldots, n_b$. 
When baskets are incorporated only in the scale parameter ($\sigma$) within a regression framework while assuming that hazards are proportional between baskets, the model can be described as follows
\begin{eqnarray}
log(\sigma_b) &=& \beta_0 + \mathbf{X}\bm{\beta} \nonumber \\
\beta_0 \sim \mbox{Normal($0, 100$)}&,& 
\beta_k \sim \mbox{Normal($0, 100$)} \nonumber \\
\kappa &\sim& \mbox{Uniform($0, 10$)}.
\end{eqnarray}
where $\mathbf{X}$ is a design matrix $\sum_{b = 1}^B n_b \times (B - 1)$ and $\bm{\beta}$ is a $(B - 1)$ vector of regression parameters, $k = 0, \ldots = (B - 1)$, $b = 1, \ldots, B$ and $i = 1, \ldots, n_b$. 


The posterior distributions for the parameters $\kappa$ and $\sigma_b$ can be easily obtained using probabilistic programming tools for Bayesian statistical modeling and inference such as JAGS \cite{Plummer2015} or STAN.

At the time of interim analysis, patients with incomplete follow-up can be considered as missing data. The missing data indicator for patient $i$ in basket $b$ is defined as 
\begin{eqnarray}
m_{bi}(c_{bi}) = \begin{cases}
1 \quad \mbox{if $t_{bi} > c_{bi}$ and $c_{bi} < T$} \\
0 \quad \mbox{if $t_{bi} \leq c_{bi}$ and $c_{bi} = T$}.
\end{cases}
\end{eqnarray}

Then, the multiple imputation is performed as follows:
\begin{enumerate}
\item Sample $(\kappa_b^{(j)}, \sigma_b^{(j)})$ from the posterior distribution for $j = 1, \ldots, J$;
\item For $j-th$ posterior sample $(\kappa_b^{(j)}, \sigma_b^{(j)})$, generate the missing responses as $y_{bi}^{(j)} | m_{bi}(c_{bi})^{(j)} = 1 \sim \mbox{Bernoulli($\omega_{bi}^{(j)}$)}$ where
\begin{eqnarray}
\omega_{bi}^{(j)} &=& P(y_{bi} = 1|m_{bi}(c_{bi})^{(j)} = 1) \nonumber\\
&=& P(t_{bi} < T|t_{bi} > x_{bi}, \kappa_b^{(j)}, \sigma_b^{(j)}) \nonumber\\
&=& \displaystyle\frac{F^{(j)}(T) - F^{(j)}(x_{bi})}{1 - F^{(j)}(x_{bi})};
\end{eqnarray}

\item Calculate the posterior distribution for each basket without information sharing using \eqref{eq:posterior_nosharing};

\item Then, calculate the level of information borrowing across baskets using \eqref{eq:weights_js} with $r_k^{(j)} = \sum_{i: m_{ki}(c_{ki}) = 0} y_{ki} + \\ \sum_{i: m_{ki}(c_{ki}) = 1} y_{ki}^{(j)}$ and $r_b^{(j)} = \sum_{i: m_{bi}(c_{bi}) = 0} y_{bi} + \sum_{i: m_{bi}(c_{bi}) = 1} y_{bi}^{(j)}$;

\item Calculate the posterior distributions with information sharing using \eqref{eq:posterior_sharing} with $r_k^{(j)} = \sum_{i: m_{ki}(c_{ki}) = 0} y_{ki} + \sum_{i: m_{ki}(c_{ki}) = 1} y_{ki}^{(j)}$;

\item Estimate the posterior probability that the drug is futile,
\begin{eqnarray}
P(\theta_b < \phi|\bm{r}, \bm{w}) = \frac{1}{J}\sum_{j = 1}^J P(\theta_b < \phi|\bm{r});
\end{eqnarray}

\item Make the decision to stop the trial for basket $b$ due to futility if $P(\theta_b < \phi|\bm{r}, \bm{w}) > \gamma_{interim}$;

\item If the trial for basket $b$ was not stopped, then the intervention is declared efficacious for basket $b$ if $P(\theta_b > \phi|\bm{r}, \bm{w}) > \gamma_{final}$ at the end of the trial.

\end{enumerate}

\section{Simulation Study}
\label{sec:simu}

A simulation study was setup to study to evaluate six strategies to handle incomplete data during interim analyses:
\begin{itemize}
\item Naive Imputation (NI): Enrolled patients that do not have an observed response at the time of interim analysis are considered as non-responders;
\item Observed data (OD): Only patients who have responded to the treatment or completed their follow-up period are considered when calculating the posterior distribution;
\item Complete data (CD): Trial accrual is suspended until responses of all enrolled patients are observed before enrolling the next patient;
\item Multiple imputation with same parameters (MI): Patients with incomplete follow-up have their responses imputed using a Weibull model with the same scale and shape parameters for all baskets;
\item Multiple imputation with covariate (MIC): Patients with incomplete follow-up have their responses imputed using a Weibull model the scale parameter modeled using a regression framework with dummy variables for each basket;
\end{itemize}

Each strategy was assessed based on the following operating characteristics calculated over 1000 simulated trials: (i) probability of early termination (PET) as the proportion of trials where the trial is stopped due futility for a specific basket; (ii) expected sample size (ESS) as the average final sample size across all simulated trials; (iii) expected trial duration (ETD) as the average time required to complete the trial across all simulated trials; (iv) expected number of correct decisions (ECD) as the average of the number of decisions of correct decisions made across all baskets in a trial, where a correct decision is defined as not rejecting a true null hypothesis (futility) or rejecting a false null hypothesis (efficacy); (v) trial-wise type I error as the proportion of trials where the null hypothesis was rejected for at least one basket that was, in reality, futile; (vi) basket-wise type I error as the proportion of trials where the null hypothesis was rejected for a specific futile basket;  and (vii) basket-wise power as the proportion of trials where the null hypothesis was rejected for a specific efficacious basket. 

\begin{figure}
    \centering
    \includegraphics[width=0.5\linewidth]{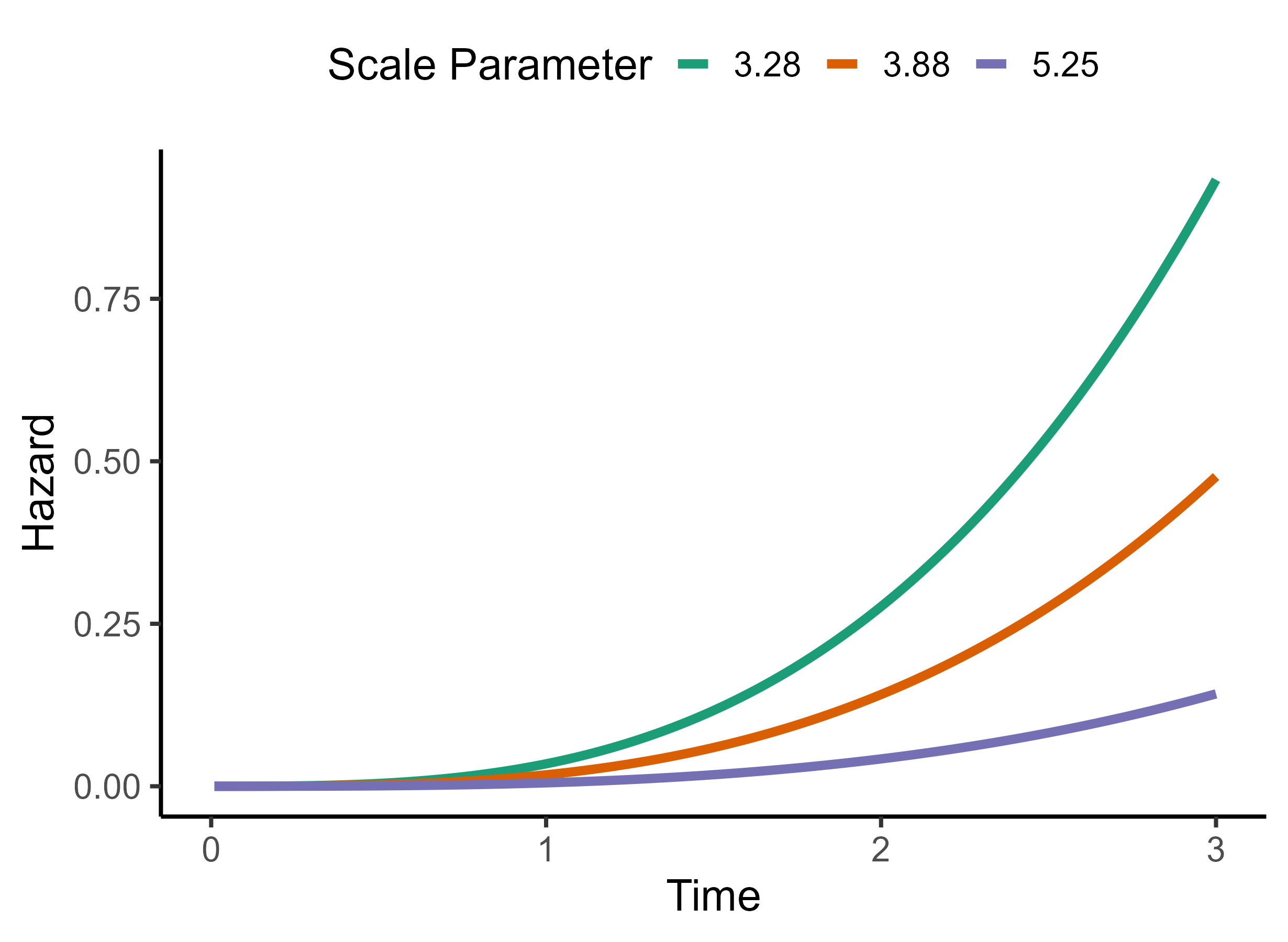}
    \caption{Scenarios for 3-month response rate: 10\% ($\sigma = 5.25$), 30\% ($\sigma = 3.88$) and 50\% ($\sigma = 3.28$) for drugs with increasing chance of response over time  $\kappa = 4$}
    \label{fig:scenarios}
\end{figure}

Figure \ref{fig:scenarios} shows the scenarios for 3-month response rate of 10\%, 30\% and 50\% for drugs with increasing chance of response over time  ($\sigma = 5.25, 3.88, 3.28$, $\kappa = 4$). We considered scenarios with 2 or 3 baskets - $B = 2, 3$ - with all  combinations of response rate resulting into a total of 16 scenarios. Our goal was to test the null hypothesis that 3-month response was equal to 30\%, $H: \phi \leq 0.3$. 

For each scenario and each basket, data was generated with patients arriving for each basket following a homogeneous Poisson process with rate $\lambda = 0.5, 1.5$ patients/month. For each basket, two interim monitoring strategies were evaluated for the basket sample sizes of 24 and 36 patients: (a) A single interim analysis was performed at the halfway point, 12 patients for a basket size of 24, and 18 patients for a basket size of 36; (b) Three interim analyzes were performed, with the first at the halfway point, and every 4 patients afterwards for a basket sample size of 24, in other words, 12, 16 and 20 patients; and every 6 patients afterwards for a basket sample size of 36, in other words, 18, 24 and 30 patients. At the time of the interim analyses for a given basket, all available data for other baskets were considered to calculate the posterior distribution. 

The prior distribution parameters were chosen as $(s_1 = 0.1, s_2 = 0.2)$ to ensure that the prior mean matches the null hypothesis. Design parameters were not optimized to reach specific power and trial-wise type I error in the benchmark CD, but the same reasonable values were chosen for all strategies, $\gamma_{interim} = 0.95$ and $\gamma_{final} = 0.975$. For the strategies that required MCMC sampling, a burn-in sample of 10000 iterations with a thinning of 70 was used with JAGS. R Code for the simulation study is available as Supplemental Material.

\FloatBarrier

\section{Results}\label{sec:results}

\begin{figure}[!h]
\centering
 \captionsetup[figure]{labelfont=bf, textfont=normalfont}
 \captionsetup[subfigure]{labelformat=parens, labelfont=bf, textfont=normalfont}

\begin{subfigure}{\textwidth}
  \centering
  \captionsetup{aboveskip=1pt, belowskip=0pt}
  \includegraphics[width=\textwidth]{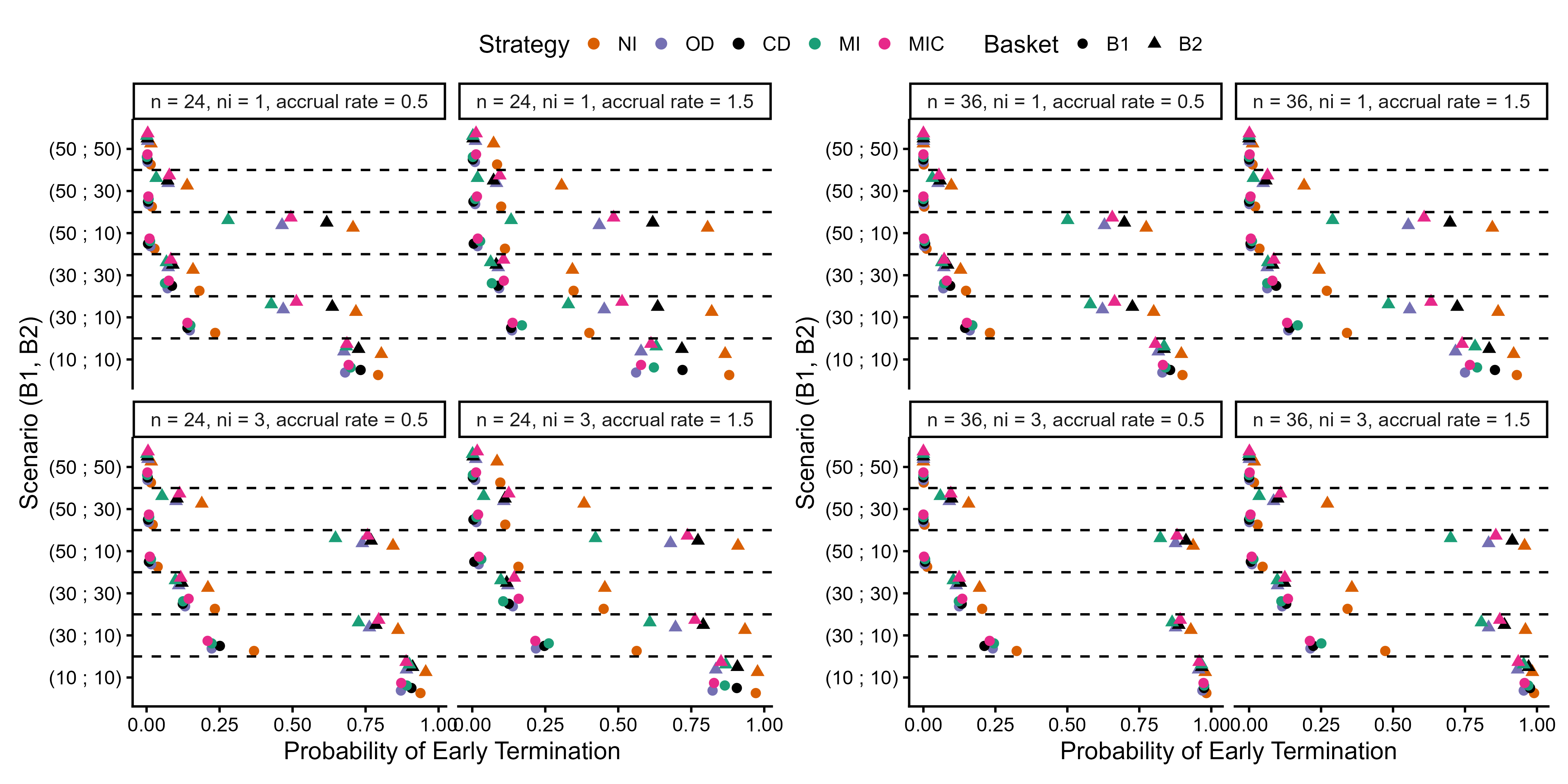} 
  \subcaption{2 baskets}\label{fig:pet_b2}
\end{subfigure} \vspace{-0.5em}

\begin{subfigure}{\textwidth}
  \centering
 \captionsetup{aboveskip=1pt, belowskip=0pt}
 \includegraphics[width=\textwidth]{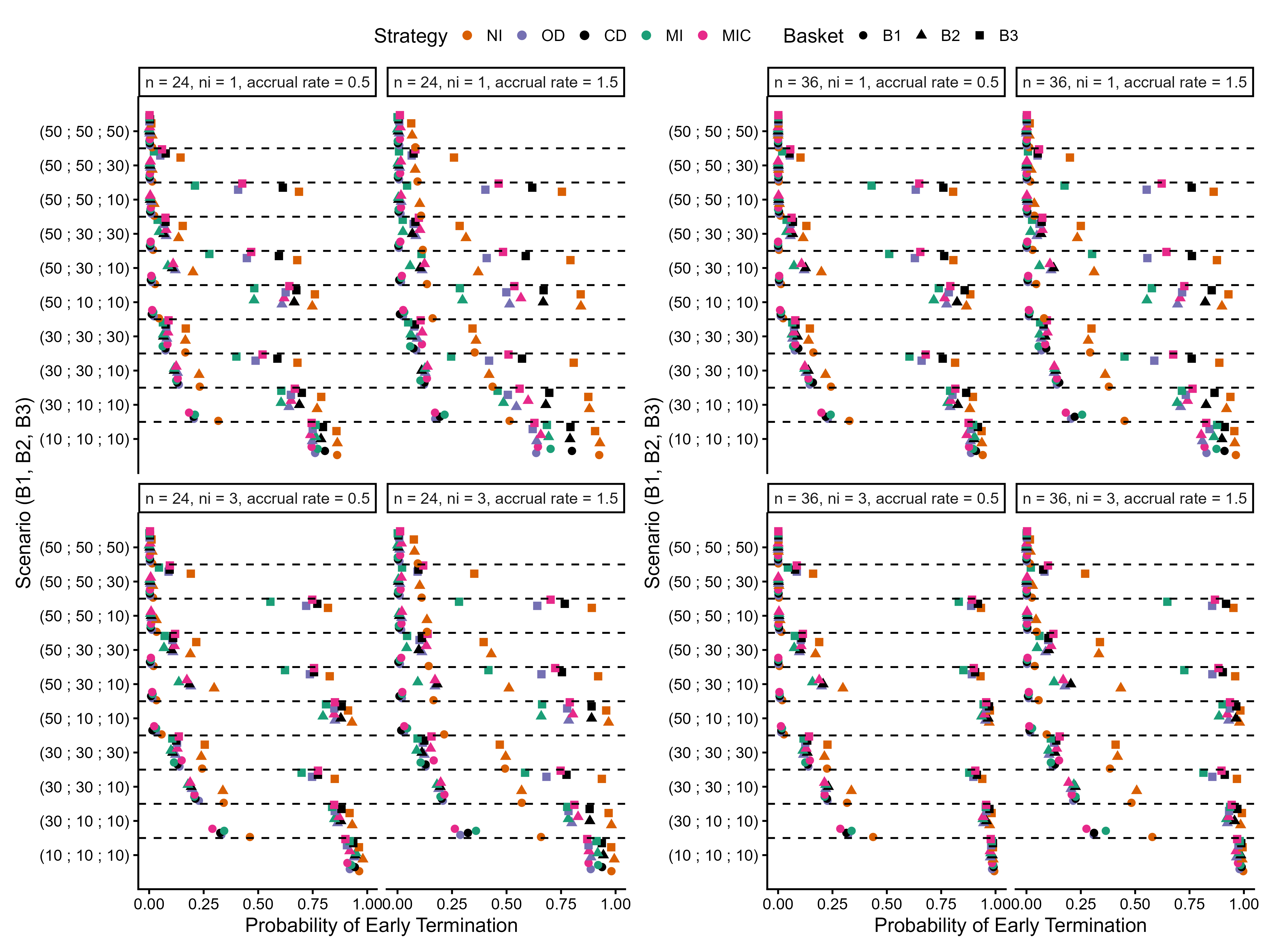} 
 \subcaption{3 baskets}\label{fig:pet_b3}

\end{subfigure} \vspace{-0.5em}

\caption{Probability of Early Termination (PET) for naïve imputation (NI), observed data (OD), complete data (CD), multiple imputation (MI), and multiple imputation with covariate (MIC) approaches  with accrual rates of 1 and 3 patients per month, 1 and 3 interim analyses, basket sample sizes of 24 and 36.}
\label{fig:pet_all}
\end{figure}

The probability of early termination is shown in Figure \ref{fig:pet_all}, and differences between candidate strategies and the approach CD as benchmark are summarized in Table \ref{tab:pet_diff}.  The PET increases with continuous monitoring, faster accrual rates, or larger sample size per basket. The NI approach consistently overestimates the PET across scenarios, including those where all baskets are efficacious, whereas the OD, MI and MIC approaches underestimate to varying degrees for futile baskets relative to the CD benchmark. The largest discrepancies between approaches occur in scenarios with at least one futile basket when accrual rate is 1.5 patients/month and a single interim analysis is conducted; in contrast, the smallest discrepancies arise when the accrual rate is 0.5 patient/month with three interim analyses. These two settings correspond to the greatest and smallest amounts of missing responses at the time of interim analyses, respectively. Increasing the sample size per basket from 24 to 36 patients reduces the magnitude of these discrepancies. When the basket response rate is 10\%, the maximum decrease in PET values ranges between -0.58 and -0.41, -0.19 and -0.09, -0.21 and -0.18 for the MI, MIC and OD approaches, respectively. When the basket response rate is 30\%, the maximum decrease in PET values ranges between -0.09 and -0.06 for MI, -0.06 and -0.01 for MIC, and -0.04 and -0.03 for OD approach. When the basket response rate is 50\%, discrepancies are negligible for all approaches, except NI with sample size of 24 patients per basket.

\begin{figure}[!h]
\centering
  \captionsetup[figure]{labelfont=bf, textfont=normalfont}
  \captionsetup[subfigure]{labelformat=parens, labelfont=bf, textfont=normalfont}

\begin{subfigure}{\textwidth}
   \centering
   \captionsetup{aboveskip=1pt, belowskip=0pt}
   \includegraphics[width=\textwidth]{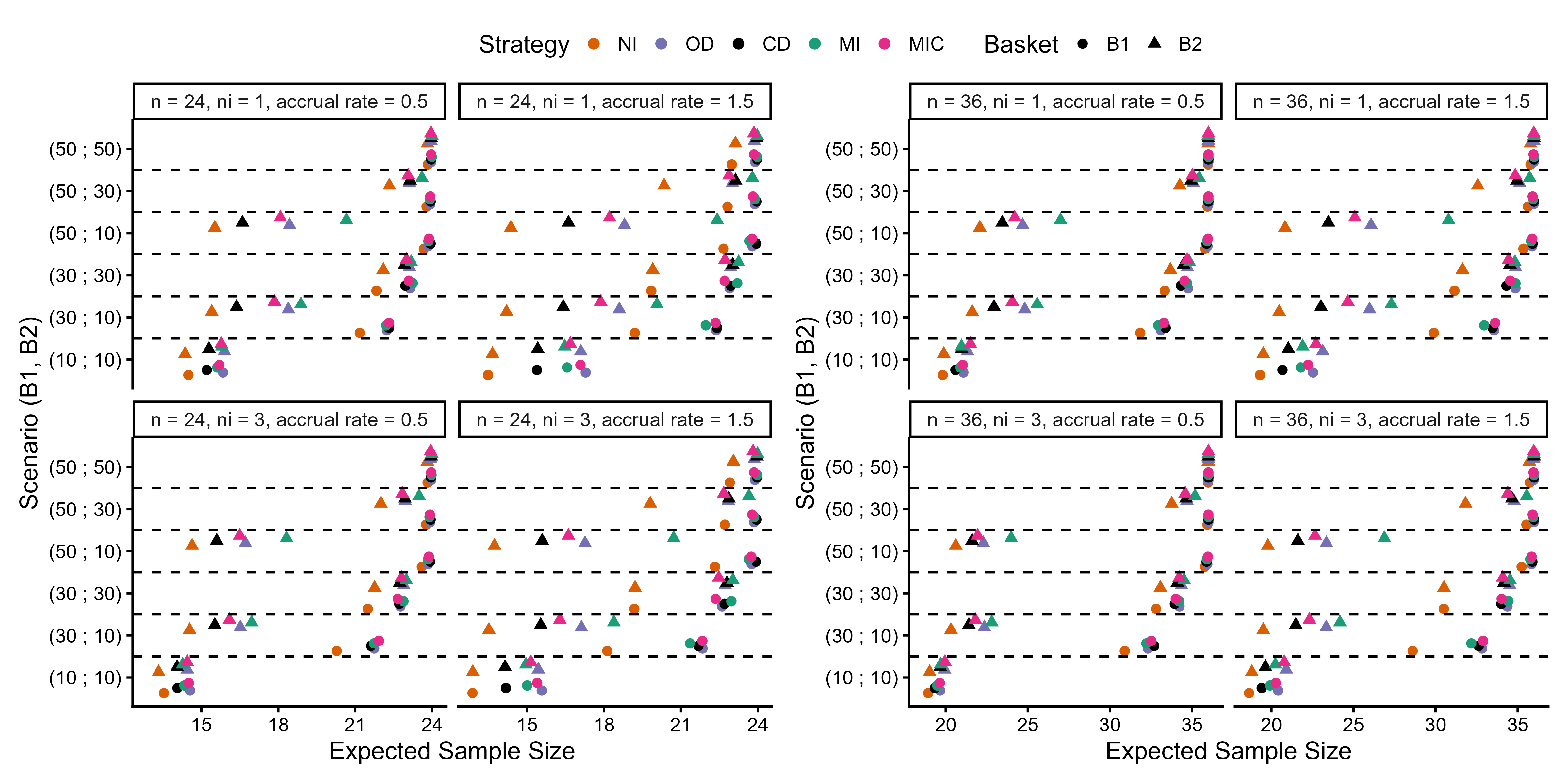} %
   \subcaption{2 baskets}\label{fig:ass_b2}
\end{subfigure} \vspace{-0.5em}

\begin{subfigure}{\textwidth}
   \centering
   \captionsetup{aboveskip=1pt, belowskip=0pt}
   \includegraphics[width=\textwidth]{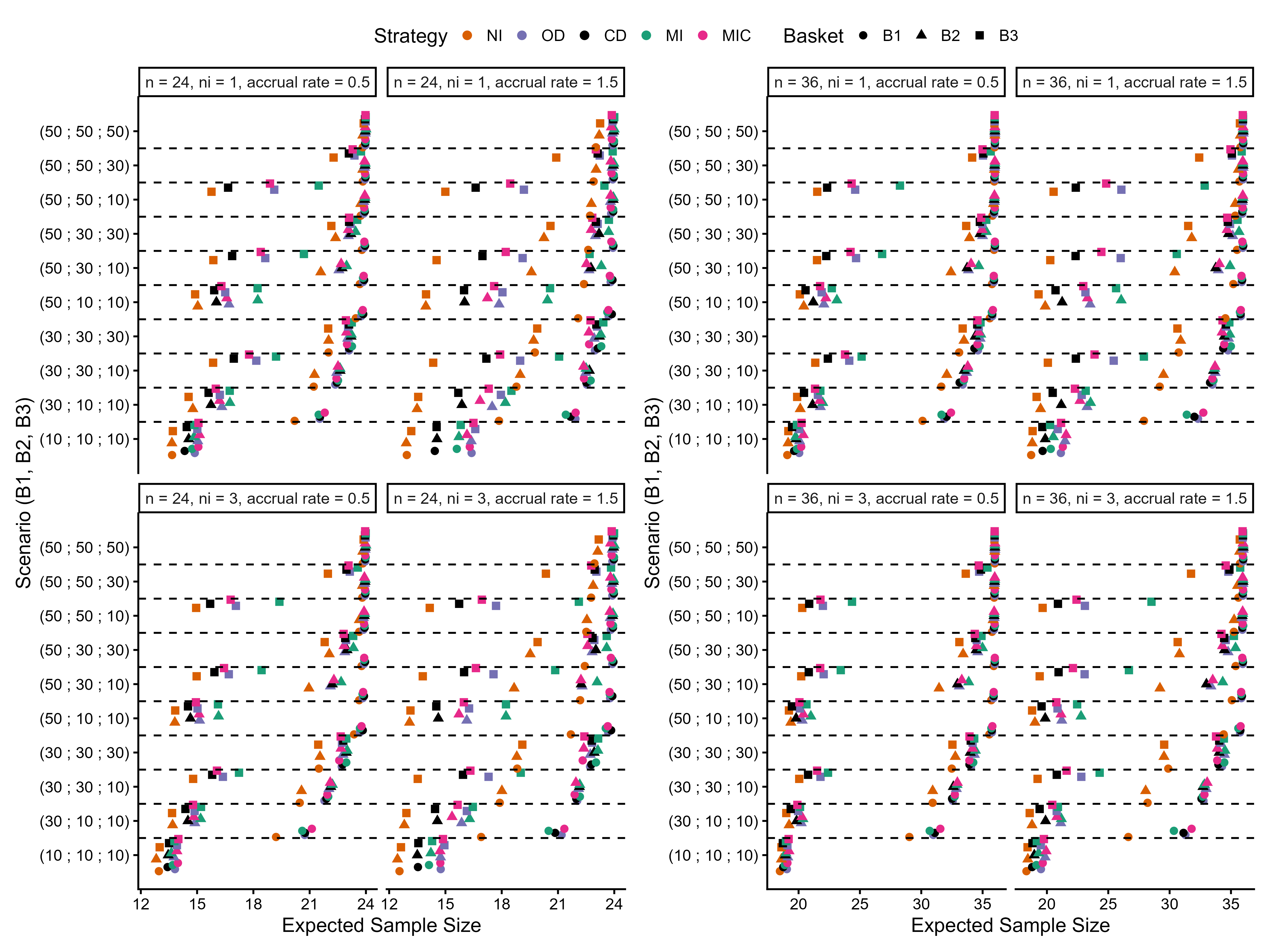} %
   \subcaption{3 baskets}\label{fig:ass_b3}
\end{subfigure} \vspace{-0.5em}

\caption{ESS for naïve imputation (NI), observed data (OD), complete data (CD), multiple imputation (MI), and multiple imputation with covariate (MIC) approaches  with accrual rates of 1 and 3 patients per month, 1 and 3 interim analyses, basket sample sizes of 24 and 36.}
\label{fig:ass_all}
\end{figure}

The ESS is presented in Figure \ref{fig:ass_all}, and percent differences relative to the sample size per basket are shown in Table \ref{tab:ass_diff}. The ESS decreases with continuous monitoring and increases with faster accrual rates. The NI approach yields the lowest ESS across all scenarios as a consequence of its highest PET values, whereas the other approaches show higher ESSs to varying extents compared to the CD benchmark. Similarly to the probability of early termination, differences between approaches are most noticeable in scenarios with at least one non-futile basket. The largest discrepancies are observed in the setting with accrual rate of 1.5 and a single interim analysis is conducted, while the smallest discrepancies occur when the accrual rate is 0.5 and three interim analyses are conducted. These discrepancies are reduced when sample size per basket increases from 24 to 36. When the basket response rate is 10\%, the maximum increase in ESS ranges between 20.3\% and 29.2\% for the MI approach,  4.7\% and 7.1\% for MIC, 8.2\% and 10.7\% for OD. When the basket response rate is 30\%, the maximum increase in sample size ranges between 2.5\% and 3.7\% for MI, 0.8\% and 2.0\% for MIC, and 0.8\% and 1.5\% for OD. For a response rate of 50\%, discrepancies remain negligible except NI with some reduction in sample size.

\begin{figure}[!h]
\centering
  \captionsetup[figure]{labelfont=bf, textfont=normalfont}
  \captionsetup[subfigure]{labelformat=parens, labelfont=bf, textfont=normalfont}

\begin{subfigure}{\textwidth}
   \centering
   \captionsetup{aboveskip=1pt, belowskip=0pt}
   \includegraphics[width=\textwidth]{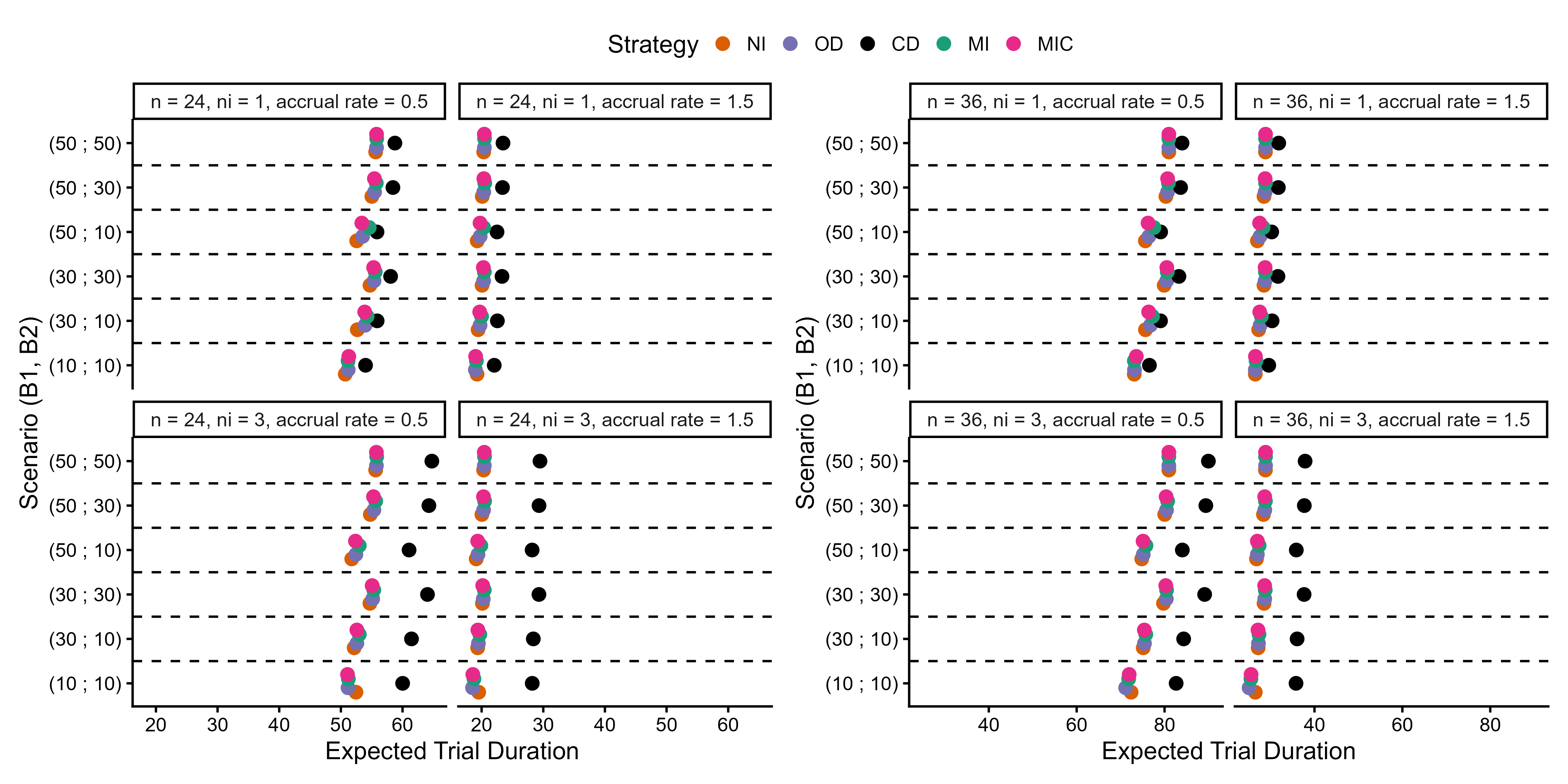} %
   \subcaption{2 baskets}\label{fig:ass_b2}
\end{subfigure} \vspace{-0.5em}

\begin{subfigure}{\textwidth}
   \centering
   \captionsetup{aboveskip=1pt, belowskip=0pt}
   \includegraphics[width=\textwidth]{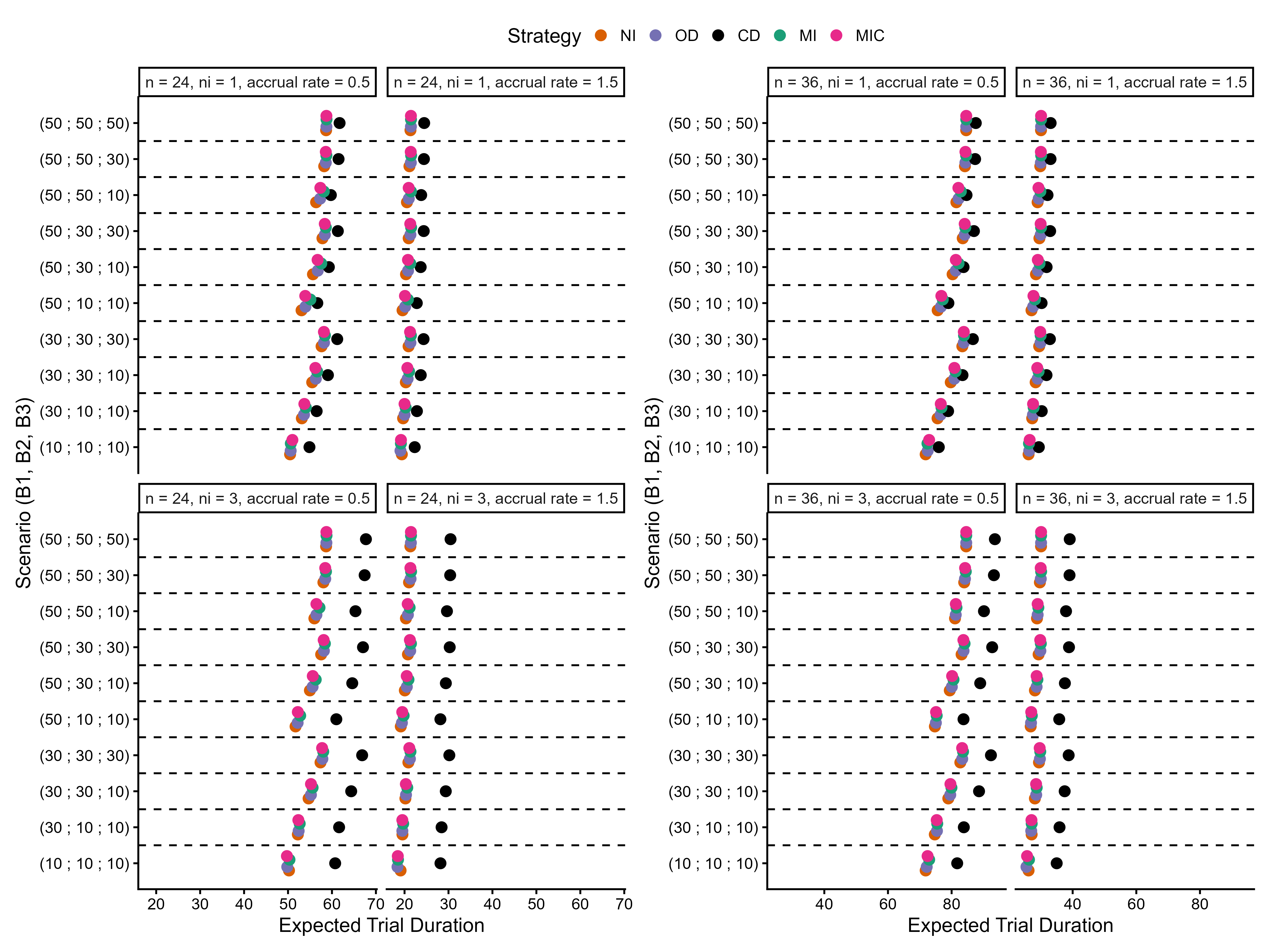} %
   \subcaption{3 baskets}\label{fig:ass_b3}
\end{subfigure} \vspace{-0.5em}

\caption{ETD for naïve imputation (NI), observed data (OD), complete data (CD), multiple imputation (MI), and multiple imputation with covariate (MIC) approaches  with accrual rates of 1 and 3 patients per month, 1 and 3 interim analyses, basket sample sizes of 24 and 36.}
\label{fig:atd_all}
\end{figure}

The ETD is displayed in Figure \ref{fig:atd_all}.   The CD approach consistently resulted in the longest expected trial duration across almost scenarios, except (10, 10), when compared to the other approaches. As expected, the difference in ETD between CD and the other approaches becomes more pronounced as the number of interim analyses increases from one to three because the other approaches remain the same while the trial duration for the CD approach increases with continuous monitoring. Furthermore, the ETD decreases with higher accrual rates, although the differences between CD and other approaches remain constant. The other operating characteristics - basket-wise type I error, trial-wise type I error, basket-wise power, ECD  - are shown in Figures \ref{fig:basketwise_alpha_all}-\ref{fig:ecd_all}. Discrepancies between approaches are negligible across scenarios as they are derived based on the final analysis that will be the same for all strategies.

\section{Discussion}
\label{sec:discussion}

We proposed continuous monitoring for basket trials and we extended the missing data framework to standard phase 2 trials to basket trials. Within this framework, we studied five strategies to handle missing data when performing interim analysis for delayed outcomes, such as objective response rate at 3 months in immunotherapy cancer trials. In our simulation study, we evaluated scenarios with 2 or 3 baskets, sample size per basket of 24 or 36 patients, accrual rate of 0.5 or 1.5 patients/month in each basket, and two interim monitoring strategies: (1) a single interim analysis at the half-way point, and (2) continuous monitoring after the half-way point. Based on our results, we recommend MIC as the most robust strategy to handle missing data achieving probability of early termination and ESS closest to the benchmark - the complete data - for fast accrual rates in heterogeneous basket scenarios, and with comparable performance to other approaches in homogeneous basket scenarios or low accrual rate. Sample size savings for MIC are observed in heterogeneous scenarios with at least one futile basket. Although these savings seem small when compared to the OD approach, their magnitude are similar to what was observed by \cite{cai_bayesian_2014}.  Furthermore, continuous monitoring in basket trials becomes feasible without significantly increasing the ETD, further increasing the savings in sample size.

Our work has some limitations: we only considered the empirical Bayesian basket trial design proposed by Fujiwara et al. \cite{fujikawa_bayesian_2020} as the computation cost becomes significantly higher with other approaches. Furthermore, we studied scenarios where all baskets had the same accrual rate, which is not a realistic assumption. However, we considered that these scenarios where the accrual is fast and slow for all baskets as limit-cases and we expect that results will fall between these two cases when heterogeneous accrual rates are considered. Lastly, we only considered the Weibull distribution to model the time to response of missing responses when an interim analysis is conducted. The Weibull distribution has the assumption that the hazard function is monotonic, which is biologically plausible as we expect a decreasing function for cytotoxic drugs  and an increasing function for immunotherapies, but other distributions that offer more flexibility might be worthy of further exploration.

In conclusion, the optimal approach to handling missing data during interim analyses depends on the specific conditions of a trial, including interim analysis schedule, accrual rate and response window. For continuous monitoring with very slow accrual, the percentage of missing data will be low at the time of each interim analysis. In such cases, high computationally demanding approaches such as MIC offer limited gain, making the OD approach as a more practical choice. However, sample sizes savings in early-phase clinical trials are always small, but these savings will be multiplied as the number of baskets increases in a given trial and an increasing number of novel drugs are tested. Lastly, the gains from the MIC approach are more significant for fast accrual rates relative to the response window. Therefore, simulation studies should be performed tailoring for a range of expected scenarios in each trial. A R-package is currently in development to support the design and evaluation of such simulation studies.   

\section*{Acknowledgments}

Research reported in this publication was supported by the National Cancer Institute Cancer Center Support Grant P30CA196521 (MAD, EM, and MM) and AIDS Malignancy Consortium Cooperative Agreement UM1CA121947 (MAD, HK, MM, DK) awarded to the Tisch Cancer Institute of the Icahn School of Medicine at Mount Sinai.

\clearpage

\begin{table}[!h]
\centering
\caption{Median (range) differences in 
         probability of early termination between 
         candidate and CD approaches
         by basket response rate across different scenarios, accrual rate, interim analysis schedule for sample sizes of 24 and 36, two and three baskets.}               
\label{tab:pet_diff}
 \small{
\begin{tabular}{c c c cccc}
  \toprule
  \multirow{2}{*}{\makecell{Sample \\ Size}} & \multirow{2}{*}{\makecell{Number of \\ Baskets}} & \multirow{2}{*}{Strategy} & \multicolumn{3}{c}{Response Rate} & \\
                          \cmidrule(lr){4-6}
                           & & & 10 & 30 & 50 \\
                            \midrule
\multirow{8}{*}{24} & \multirow{4}{*}{2} & NI & 0.08 (0.03 ; 0.19) & 0.17 (0.07 ; 0.34) & 0.05 (0.01 ; 0.15) \\ 
   \cmidrule(lr){3-6}
                            &  & OD & -0.09 (-0.18 ; -0.02) & 0.00 (-0.03 ; 0.01) & 0.01 (0.00 ; 0.01) \\ 
   \cmidrule(lr){3-6}
                            &  & MI & -0.09 (-0.48 ; -0.01) & -0.02 (-0.07 ; 0.04) & 0.00 (0.00 ; 0.02) \\ 
   \cmidrule(lr){3-6}
                            &  & MIC & -0.05 (-0.14 ; 0.01) & 0.01 (-0.04 ; 0.03) & 0.01 (0.00 ; 0.02) \\ 
   \cmidrule(lr){2-6}
                            & \multirow{4}{*}{3} & NI & 0.08 (0.02 ; 0.24) & 0.16 (0.07 ; 0.38) & 0.05 (0.01 ; 0.20) \\ 
   \cmidrule(lr){3-6}
                            &  & OD & -0.07 (-0.21 ; -0.01) & 0.00 (-0.04 ; 0.02) & 0.00 (0.00 ; 0.02) \\ 
   \cmidrule(lr){3-6}
                            &  & MI & -0.11 (-0.57 ; 0.00) & -0.02 (-0.09 ; 0.04) & 0.00 (0.00 ; 0.03) \\ 
   \cmidrule(lr){3-6}
                            &  & MIC & -0.06 (-0.19 ; 0.00) & 0.01 (-0.06 ; 0.05) & 0.01 (0.00 ; 0.01) \\ 
   \midrule
                           \multirow{8}{*}{36} & \multirow{4}{*}{2} & NI & 0.05 (0.01 ; 0.15) & 0.12 (0.04 ; 0.25) & 0.01 (0.00 ; 0.04) \\ 
   \cmidrule(lr){3-6}
                            &  & OD & -0.05 (-0.16 ; -0.01) & -0.01 (-0.03 ; 0.03) & 0.00 (0.00 ; 0.00) \\ 
   \cmidrule(lr){3-6}
                            &  & MI & -0.06 (-0.41 ; 0.00) & -0.02 (-0.06 ; 0.03) & 0.00 (0.00 ; 0.01) \\ 
   \cmidrule(lr){3-6}
                            &  & MIC & -0.03 (-0.09 ; 0.01) & 0.00 (-0.01 ; 0.02) & 0.00 (0.00 ; 0.00) \\ 
   \cmidrule(lr){2-6}
                            & \multirow{4}{*}{3} & NI & 0.03 (0.00 ; 0.13) & 0.14 (0.05 ; 0.29) & 0.01 (0.00 ; 0.08) \\ 
   \cmidrule(lr){3-6}
                            &  & OD & -0.04 (-0.21 ; -0.01) & -0.01 (-0.03 ; 0.01) & 0.00 (0.00 ; 0.00) \\ 
   \cmidrule(lr){3-6}
                            &  & MI & -0.05 (-0.58 ; 0.00) & -0.02 (-0.08 ; 0.05) & 0.00 (0.00 ; 0.01) \\ 
   \cmidrule(lr){3-6}
                            &  & MIC & -0.03 (-0.14 ; -0.01) & -0.01 (-0.04 ; 0.02) & 0.00 (0.00 ; 0.00) \\ 
   \bottomrule
\end{tabular}
}
\end{table}

\begin{table}[h]
\centering
\caption{Median (range) percent differences in 
         ESS between 
         candidate and CD approaches
         by basket response rate across different scenarios, accrual rate, interim analysis schedule for sample sizes of 24 and 36, two and three baskets.}               
\label{tab:ass_diff}
 \small{
\begin{tabular}{c c c cccc}
  \toprule
  \multirow{2}{*}{\makecell{Sample \\ Size}} & \multirow{2}{*}{\makecell{Number of \\ Baskets}} & \multirow{2}{*}{Strategy} & \multicolumn{3}{c}{Response Rate} & \\
                          \cmidrule(lr){4-6}
                           & & & 10 & 30 & 50 \\
                            \midrule
\multirow{8}{*}{24} & \multirow{4}{*}{2} & NI & -4.85 (-9.40 ; -2.20) & -8.60 (-14.90 ; -3.30) & -2.45 (-6.70 ; -0.50) \\ 
   \cmidrule(lr){3-6}
                            &  & OD & 6.20 (1.70 ; 9.10) & -0.10 (-0.50 ; 0.80) & -0.35 (-0.70 ; 0.00) \\ 
   \cmidrule(lr){3-6}
                            &  & MI & 5.45 (0.90 ; 24.10) & 1.10 (-1.80 ; 3.30) & -0.05 (-1.10 ; 0.10) \\ 
   \cmidrule(lr){3-6}
                            &  & MIC & 4.25 (1.60 ; 7.10) & -0.30 (-1.50 ; 1.30) & -0.40 (-0.80 ; 0.00) \\ 
   \cmidrule(lr){2-6}
                            & \multirow{4}{*}{3} & NI & -4.30 (-11.80 ; -1.90) & -7.85 (-17.20 ; -3.40) & -2.40 (-8.90 ; -0.30) \\ 
   \cmidrule(lr){3-6}
                            &  & OD & 5.55 (1.30 ; 10.70) & -0.20 (-1.20 ; 1.40) & -0.30 (-0.70 ; 0.10) \\ 
   \cmidrule(lr){3-6}
                            &  & MI & 7.80 (0.70 ; 28.60) & 0.90 (-1.40 ; 3.60) & 0.00 (-1.20 ; 0.20) \\ 
   \cmidrule(lr){3-6}
                            &  & MIC & 3.35 (1.10 ; 9.30) & -0.60 (-2.30 ; 2.00) & -0.35 (-0.70 ; 0.10) \\ 
   \midrule
                           \multirow{8}{*}{36} & \multirow{4}{*}{2} & NI & -3.35 (-7.30 ; -1.10) & -5.80 (-11.20 ; -1.90) & -0.45 (-1.80 ; 0.00) \\ 
   \cmidrule(lr){3-6}
                            &  & OD & 3.45 (0.70 ; 8.20) & 0.65 (-1.10 ; 1.50) & 0.00 (-0.10 ; 0.10) \\ 
   \cmidrule(lr){3-6}
                            &  & MI & 3.50 (0.00 ; 20.30) & 1.00 (-1.40 ; 2.50) & 0.00 (-0.40 ; 0.10) \\ 
   \cmidrule(lr){3-6}
                            &  & MIC & 2.35 (0.80 ; 4.70) & 0.15 (-0.70 ; 0.80) & 0.00 (-0.30 ; 0.20) \\ 
   \cmidrule(lr){2-6}
                            & \multirow{4}{*}{3} & NI & -2.15 (-6.30 ; -0.30) & -6.60 (-13.00 ; -2.40) & -0.65 (-3.80 ; 0.00) \\ 
   \cmidrule(lr){3-6}
                            &  & OD & 3.45 (0.50 ; 10.30) & 0.45 (-0.50 ; 1.20) & 0.00 (-0.10 ; 0.10) \\ 
   \cmidrule(lr){3-6}
                            &  & MI & 4.30 (-0.10 ; 29.20) & 1.20 (-2.20 ; 3.70) & 0.00 (-0.50 ; 0.20) \\ 
   \cmidrule(lr){3-6}
                            &  & MIC & 2.75 (0.80 ; 6.90) & 0.30 (-0.80 ; 2.00) & 0.00 (-0.20 ; 0.20) \\ 
   \bottomrule
\end{tabular}
}
\end{table}

\clearpage

\bibliographystyle{ama}
\bibliography{references}

\clearpage

\appendix

\renewcommand{\thesection}{S\arabic{section}}

\renewcommand{\thefigure}{S\arabic{figure}}
\setcounter{figure}{0}
\renewcommand{\thetable}{S\arabic{table}}
\setcounter{table}{0}

\section{Supplementary Material - Figures}

\begin{figure}[!h]
\centering
  \captionsetup[figure]{labelfont=bf, textfont=normalfont}
  \captionsetup[subfigure]{labelformat=parens, labelfont=bf, textfont=normalfont}

\begin{subfigure}{\textwidth}
   \centering
   \captionsetup{aboveskip=1pt, belowskip=0pt}
   \includegraphics[width=\textwidth]{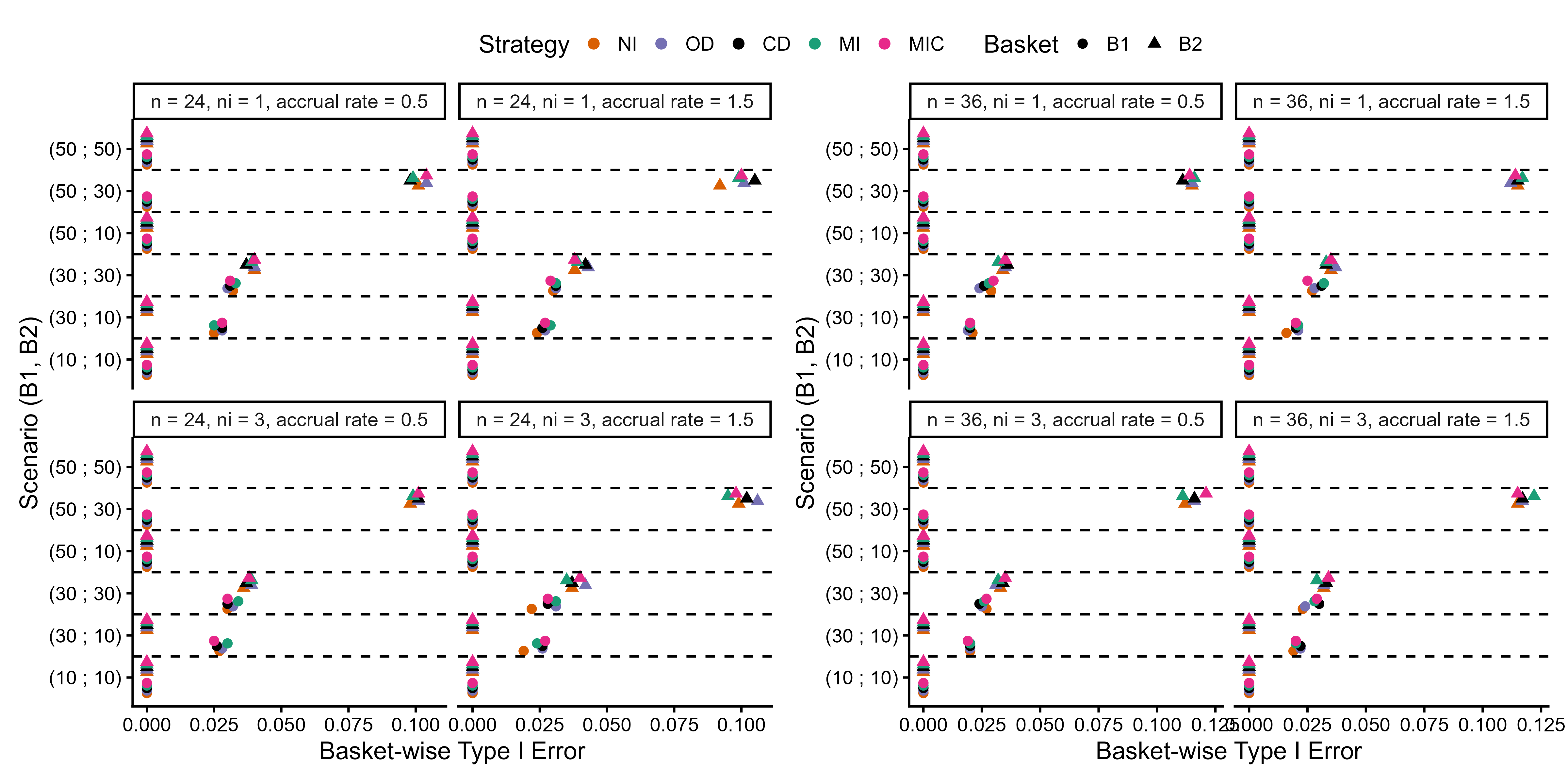} %
   \subcaption{2 baskets}\label{fig:basketwise_alpha_b2}
\end{subfigure} \vspace{-0.5em}

\begin{subfigure}{\textwidth}
   \centering
   \captionsetup{aboveskip=1pt, belowskip=0pt}
   \includegraphics[width=\textwidth]{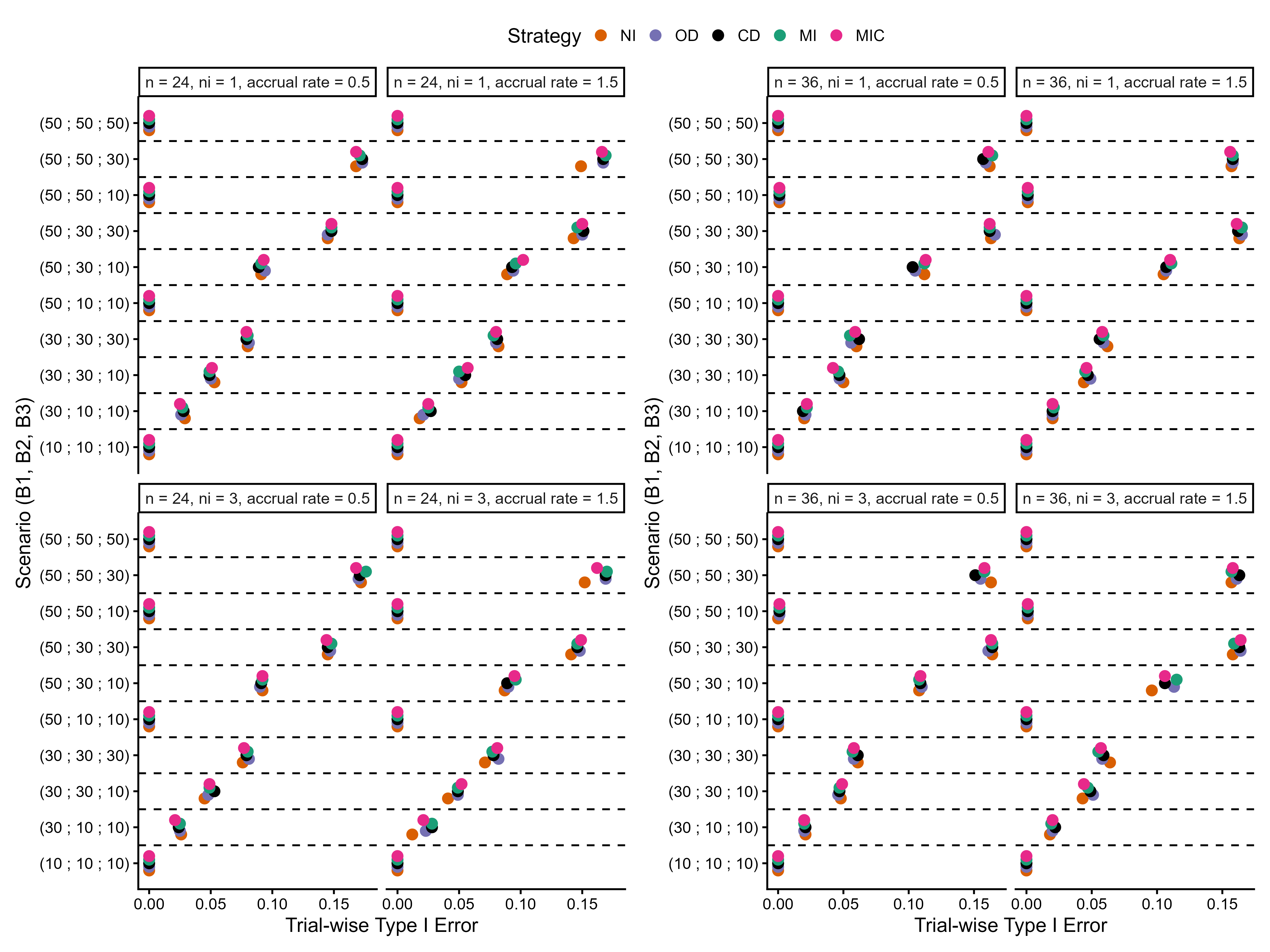} %
   \subcaption{3 baskets}\label{fig:basketwise_alpha_b3}
\end{subfigure} \vspace{-0.5em}

\caption{Basket-wise type I error for naïve imputation (NI), observed data (OD), complete data (CD), multiple imputation (MI), and multiple imputation with covariate (MIC) approaches  with accrual rates of 1 and 3 patients per month, 1 and 3 interim analyses, basket sample sizes of 24 and 36.}
\label{fig:basketwise_alpha_all}
\end{figure}

\begin{figure}[!h]
\centering
  \captionsetup[figure]{labelfont=bf, textfont=normalfont}
  \captionsetup[subfigure]{labelformat=parens, labelfont=bf, textfont=normalfont}

\begin{subfigure}{\textwidth}
   \centering
   \captionsetup{aboveskip=1pt, belowskip=0pt}
   \includegraphics[width=\textwidth]{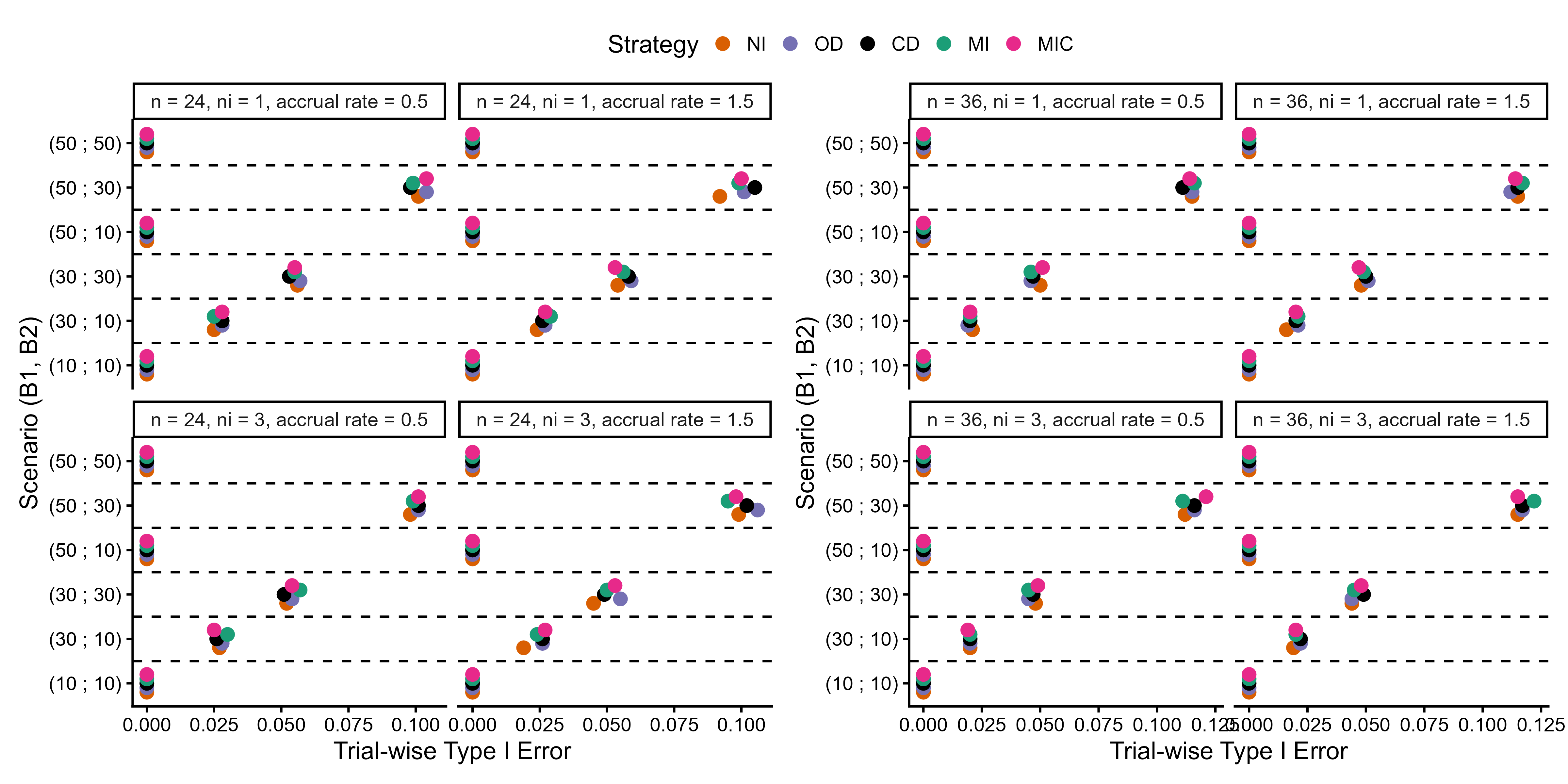} %
   \subcaption{2 baskets}\label{fig:trialwise_alpha_b2}
\end{subfigure} \vspace{-0.5em}

\begin{subfigure}{\textwidth}
   \centering
   \captionsetup{aboveskip=1pt, belowskip=0pt}
   \includegraphics[width=\textwidth]{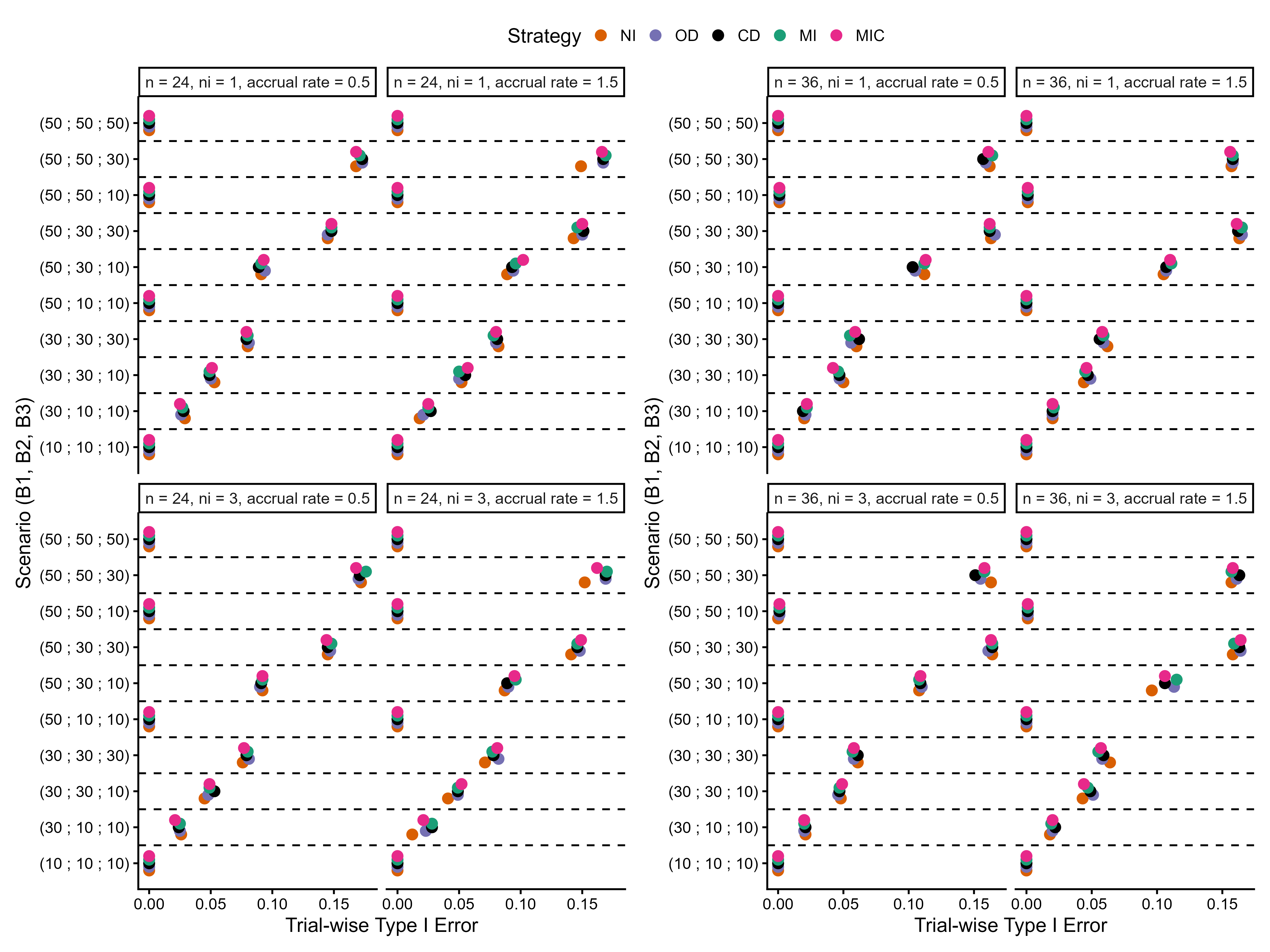} %
   \subcaption{3 baskets}\label{fig:trialwise_alpha_b3}
\end{subfigure} \vspace{-0.5em}

\caption{Trial-wise type I error for naïve imputation (NI), observed data (OD), complete data (CD), multiple imputation (MI), and multiple imputation with covariate (MIC) approaches  with accrual rates of 1 and 3 patients per month, 1 and 3 interim analyses, basket sample sizes of 24 and 36.}
\label{fig:trialwise_alpha_all}
\end{figure}

\begin{figure}[!h]
\centering
  \captionsetup[figure]{labelfont=bf, textfont=normalfont}
  \captionsetup[subfigure]{labelformat=parens, labelfont=bf, textfont=normalfont}

\begin{subfigure}{\textwidth}
   \centering
   \captionsetup{aboveskip=1pt, belowskip=0pt}
   \includegraphics[width=\textwidth]{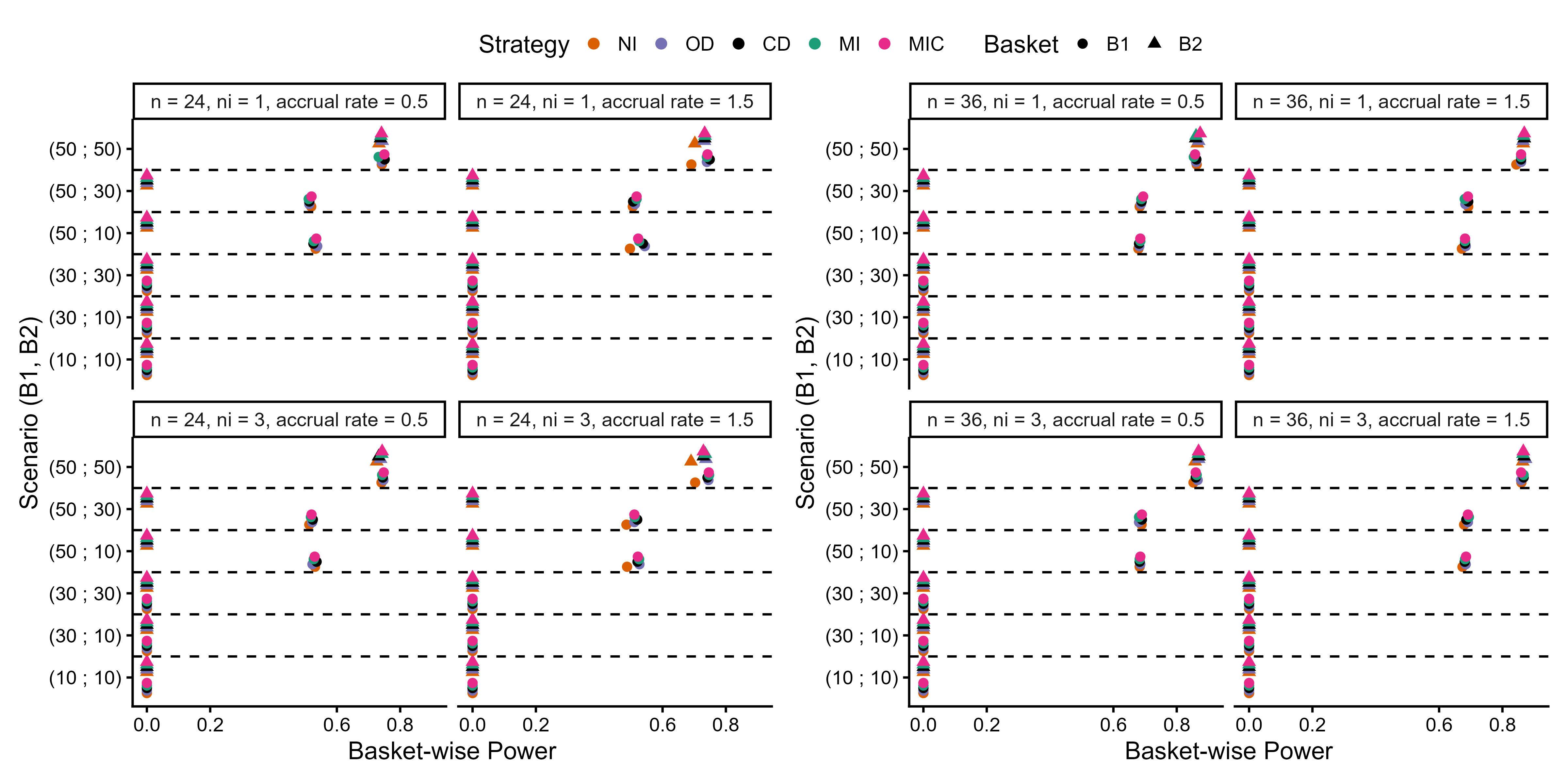} %
   \subcaption{2 baskets}\label{fig:basketwise_power_b2}
\end{subfigure} \vspace{-0.5em}

\begin{subfigure}{\textwidth}
   \centering
   \captionsetup{aboveskip=1pt, belowskip=0pt}
   \includegraphics[width=\textwidth]{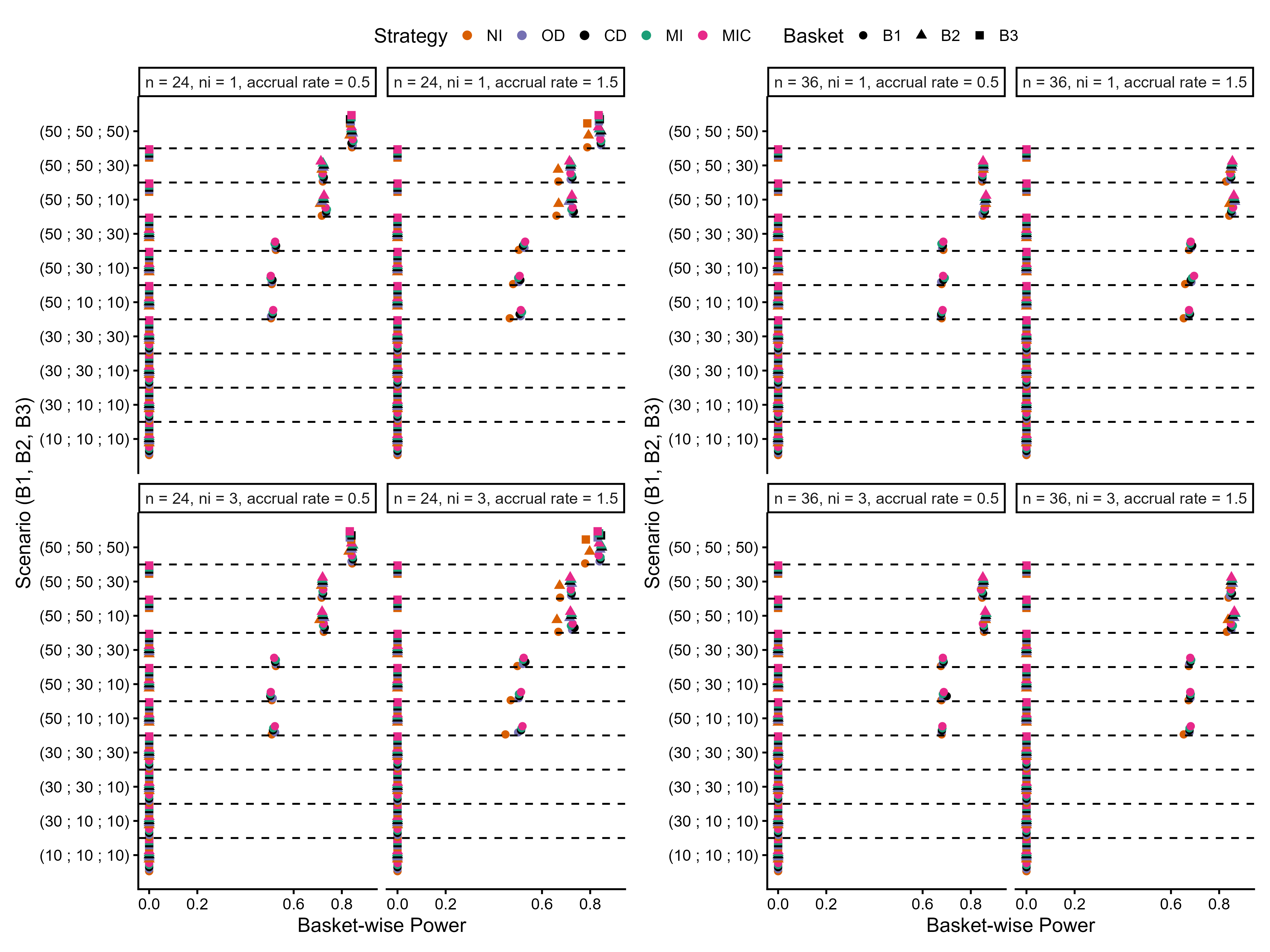} %
   \subcaption{3 baskets}\label{fig:basketwise_power_b3}
\end{subfigure} \vspace{-0.5em}

\caption{Basket-wise power for naïve imputation (NI), observed data (OD), complete data (CD), multiple imputation (MI), and multiple imputation with covariate (MIC) approaches  with accrual rates of 1 and 3 patients per month, 1 and 3 interim analyses, basket sample sizes of 24 and 36.}
\label{fig:basketwise_power_all}
\end{figure}

\begin{figure}[!h]
\centering
  \captionsetup[figure]{labelfont=bf, textfont=normalfont}
  \captionsetup[subfigure]{labelformat=parens, labelfont=bf, textfont=normalfont}

\begin{subfigure}{\textwidth}
   \centering
   \captionsetup{aboveskip=1pt, belowskip=0pt}
   \includegraphics[width=\textwidth]{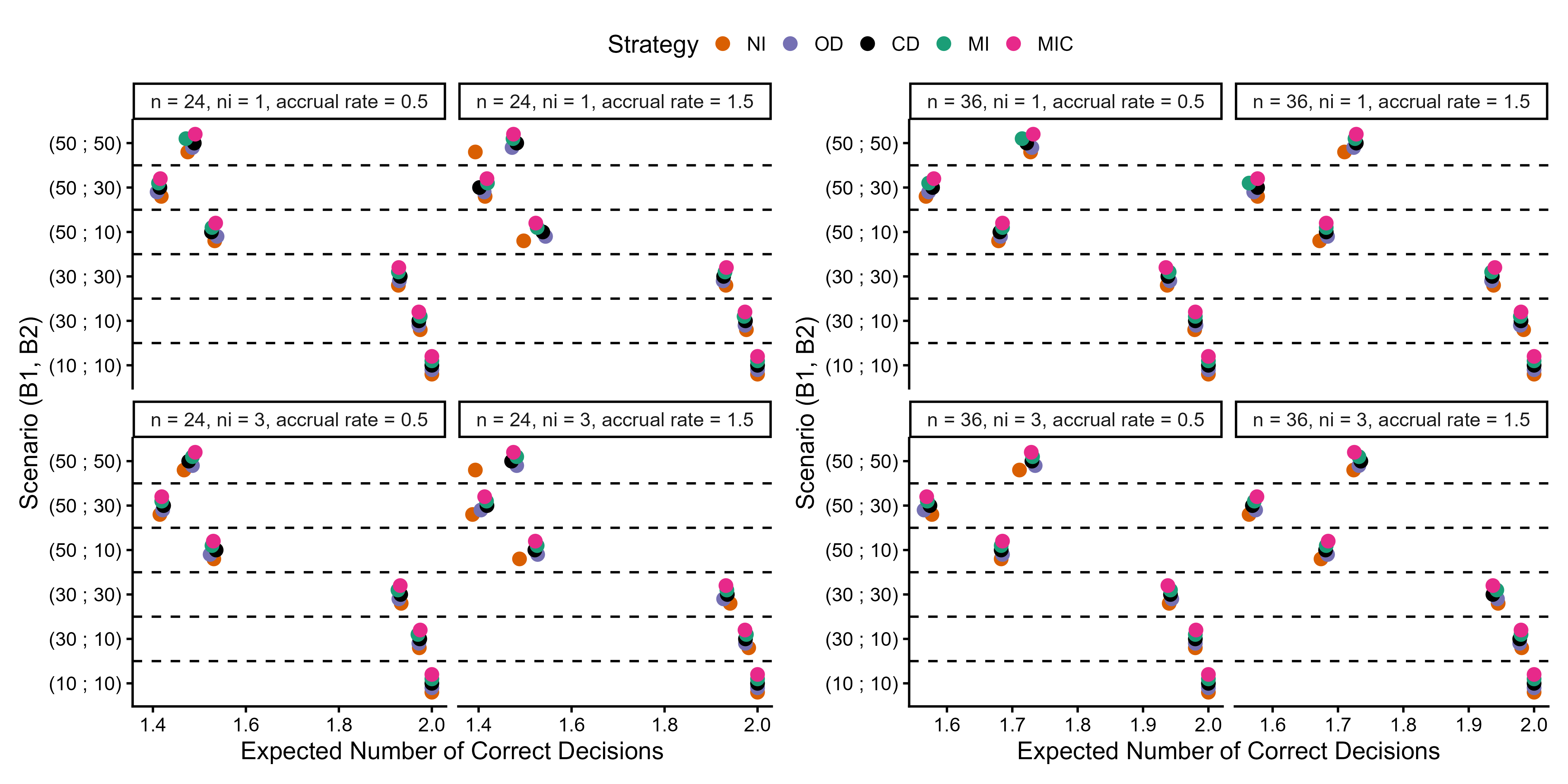} %
   \subcaption{2 baskets}\label{fig:ecd_b2}
\end{subfigure} \vspace{-0.5em}

\begin{subfigure}{\textwidth}
   \centering
   \captionsetup{aboveskip=1pt, belowskip=0pt}
   \includegraphics[width=\textwidth]{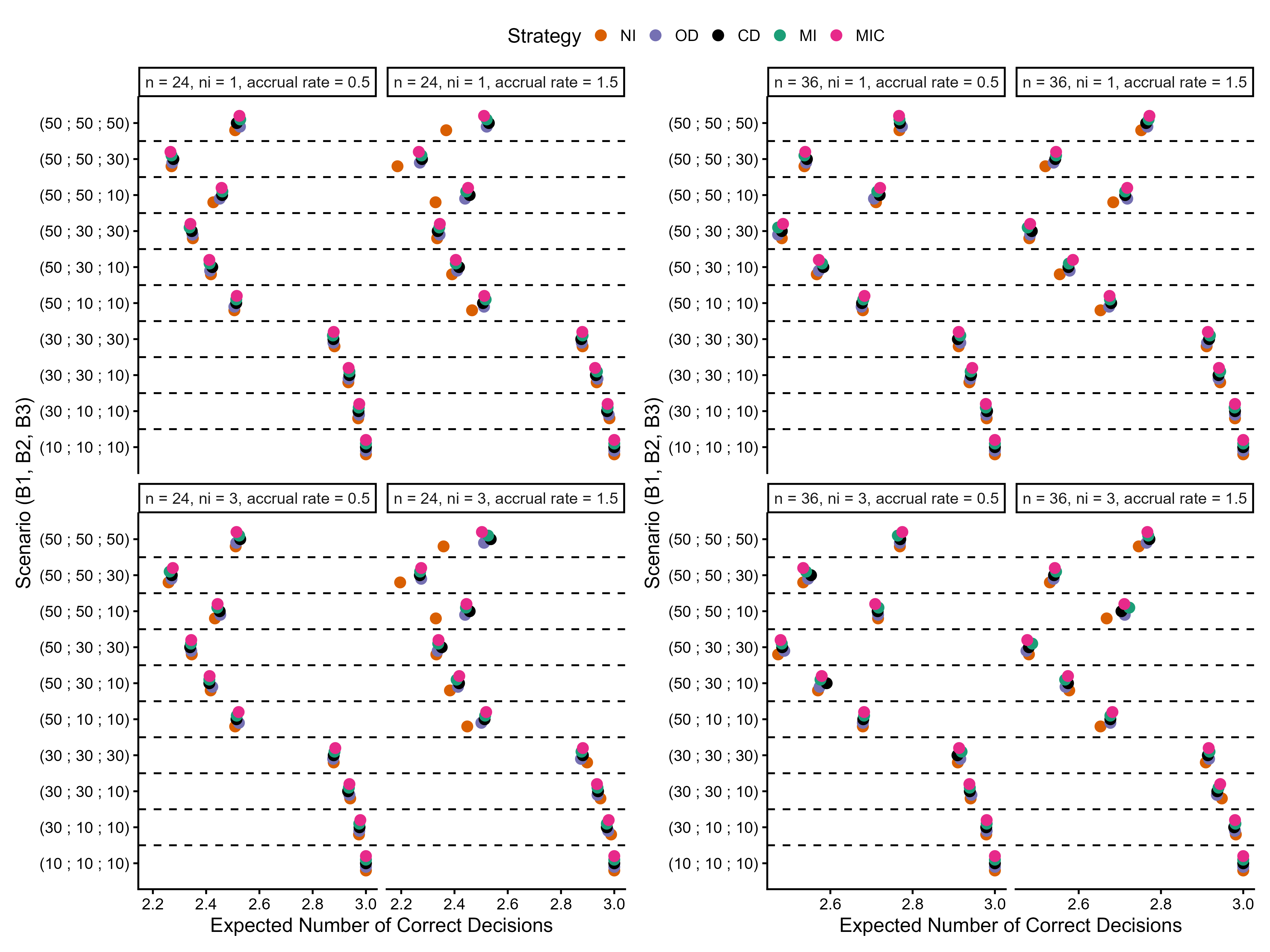} %
   \subcaption{3 baskets}\label{fig:ecd_b3}
\end{subfigure} \vspace{-0.5em}

\caption{Expected number of correct decisions for naïve imputation (NI), observed data (OD), complete data (CD), multiple imputation (MI), and multiple imputation with covariate (MIC) approaches  with accrual rates of 1 and 3 patients per month, 1 and 3 interim analyses, basket sample sizes of 24 and 36.}
\label{fig:ecd_all}
\end{figure}

\clearpage

\section{Supplementary Material - Tables}

\captionsetup[table]{labelfont=bf, labelsep=colon}

\begin{center}
\small
\renewcommand{\arraystretch}{0.8}
\begin{tabularx}{\textwidth}{c c c *{10}{X}}
\caption{Probability of Early Termination (PET) under varying response rates, 
              two accrual rates, two interim analysis scenarios, 
              and sample sizes of 24 and 36 for two baskets.} \label{tab:pet2}\\
\toprule
\multirow{2}{*}{\makecell{\textbf{Accrual} \\ \textbf{Rate}}} &
\multirow{2}{*}{\makecell{\textbf{Response} \\ \textbf{Rate (B1, B2)}}} &
\multirow{2}{*}{\textbf{Basket}} &
\multicolumn{5}{c}{\textbf{\# interim analyses = 1}} & \multicolumn{5}{c}{\textbf{\# interim analyses = 3}}
\\
\cmidrule(lr){4-8}\cmidrule(lr){9-13}
 &  &  & 
\textbf{NI} & \textbf{OD} & \textbf{CD} & \textbf{MI} & \textbf{MIC} & \textbf{NI} & \textbf{OD} & \textbf{CD} & \textbf{MI} & \textbf{MIC}
\endfirsthead

\toprule
\multirow{2}{*}{\makecell{\textbf{Accrual}\\\textbf{Rate}}} &
\multirow{2}{*}{\makecell{\textbf{Response}\\\textbf{Rate (B1, B2)}}} &
\multirow{2}{*}{\textbf{Basket}} &
\multicolumn{5}{c}{\textbf{\# interim analyses = 1}} & \multicolumn{5}{c}{\textbf{\# interim analyses = 3}}
\\
\cmidrule(lr){4-8}\cmidrule(lr){9-13}
 &  &  & 
\textbf{NI} & \textbf{OD} & \textbf{CD} & \textbf{MI} & \textbf{MIC} & \textbf{NI} & \textbf{OD} & \textbf{CD} & \textbf{MI} & \textbf{MIC}
\\
\midrule
\endhead

\multicolumn{13}{r}{\small\textit{Continued on next page}}\\
\hline
\endfoot

\bottomrule
\endlastfoot
\bottomrule
\endlastfoot
\hline
\midrule 
\multicolumn{13}{c}{\textcolor{red}{\textbf{Sample size = 24}}} \\ 
  \midrule
\multirow{12}{*}{0.5} & \multirow{2}{*}{(10 ; 10)} & B1 & 0.79 & 0.68 & 0.73 & 0.70 & 0.69 & 0.94 & 0.87 & 0.91 & 0.89 & 0.87 \\ 
   &  & B2 & 0.80 & 0.68 & 0.73 & 0.68 & 0.69 & 0.96 & 0.89 & 0.91 & 0.90 & 0.89 \\ 
   \cmidrule{2-13}
 & \multirow{2}{*}{(30 ; 10)} & B1 & 0.23 & 0.15 & 0.14 & 0.15 & 0.14 & 0.37 & 0.22 & 0.25 & 0.22 & 0.21 \\ 
   &  & B2 & 0.72 & 0.47 & 0.64 & 0.43 & 0.51 & 0.86 & 0.76 & 0.78 & 0.73 & 0.80 \\ 
   \cmidrule{2-13}
 & \multirow{2}{*}{(30 ; 30)} & B1 & 0.18 & 0.07 & 0.09 & 0.06 & 0.08 & 0.23 & 0.13 & 0.12 & 0.12 & 0.14 \\ 
   &  & B2 & 0.16 & 0.07 & 0.09 & 0.07 & 0.08 & 0.21 & 0.11 & 0.12 & 0.10 & 0.12 \\ 
   \cmidrule{2-13}
 & \multirow{2}{*}{(50 ; 10)} & B1 & 0.03 & 0.01 & 0.00 & 0.01 & 0.01 & 0.04 & 0.02 & 0.01 & 0.01 & 0.01 \\ 
   &  & B2 & 0.71 & 0.46 & 0.62 & 0.28 & 0.49 & 0.84 & 0.74 & 0.77 & 0.65 & 0.76 \\ 
   \cmidrule{2-13}
 & \multirow{2}{*}{(50 ; 30)} & B1 & 0.02 & 0.01 & 0.00 & 0.01 & 0.01 & 0.02 & 0.01 & 0.00 & 0.01 & 0.01 \\ 
   &  & B2 & 0.14 & 0.07 & 0.07 & 0.03 & 0.08 & 0.19 & 0.10 & 0.10 & 0.05 & 0.11 \\ 
   \cmidrule{2-13}
 & \multirow{2}{*}{(50 ; 50)} & B1 & 0.01 & 0.00 & 0.00 & 0.00 & 0.00 & 0.01 & 0.00 & 0.00 & 0.00 & 0.00 \\ 
   &  & B2 & 0.01 & 0.00 & 0.00 & 0.00 & 0.00 & 0.02 & 0.00 & 0.00 & 0.00 & 0.00 \\ 
   \midrule
\multirow{12}{*}{1.5} & \multirow{2}{*}{(10 ; 10)} & B1 & 0.88 & 0.56 & 0.72 & 0.62 & 0.58 & 0.97 & 0.82 & 0.91 & 0.86 & 0.83 \\ 
   &  & B2 & 0.87 & 0.58 & 0.72 & 0.63 & 0.61 & 0.98 & 0.83 & 0.91 & 0.87 & 0.85 \\ 
   \cmidrule{2-13}
 & \multirow{2}{*}{(30 ; 10)} & B1 & 0.40 & 0.14 & 0.13 & 0.17 & 0.14 & 0.56 & 0.22 & 0.25 & 0.26 & 0.22 \\ 
   &  & B2 & 0.82 & 0.45 & 0.64 & 0.33 & 0.51 & 0.93 & 0.70 & 0.79 & 0.61 & 0.76 \\ 
   \cmidrule{2-13}
 & \multirow{2}{*}{(30 ; 30)} & B1 & 0.35 & 0.09 & 0.09 & 0.07 & 0.11 & 0.45 & 0.14 & 0.12 & 0.11 & 0.16 \\ 
   &  & B2 & 0.34 & 0.09 & 0.08 & 0.06 & 0.11 & 0.45 & 0.12 & 0.12 & 0.10 & 0.14 \\ 
   \cmidrule{2-13}
 & \multirow{2}{*}{(50 ; 10)} & B1 & 0.11 & 0.02 & 0.00 & 0.03 & 0.02 & 0.16 & 0.02 & 0.01 & 0.03 & 0.02 \\ 
   &  & B2 & 0.81 & 0.43 & 0.62 & 0.13 & 0.48 & 0.91 & 0.68 & 0.77 & 0.42 & 0.74 \\ 
   \cmidrule{2-13}
 & \multirow{2}{*}{(50 ; 30)} & B1 & 0.10 & 0.01 & 0.00 & 0.01 & 0.02 & 0.11 & 0.01 & 0.00 & 0.01 & 0.02 \\ 
   &  & B2 & 0.31 & 0.08 & 0.07 & 0.02 & 0.09 & 0.38 & 0.11 & 0.11 & 0.04 & 0.12 \\ 
   \cmidrule{2-13}
 & \multirow{2}{*}{(50 ; 50)} & B1 & 0.08 & 0.01 & 0.00 & 0.00 & 0.01 & 0.10 & 0.01 & 0.00 & 0.00 & 0.01 \\ 
   &  & B2 & 0.07 & 0.01 & 0.00 & 0.00 & 0.01 & 0.08 & 0.01 & 0.00 & 0.00 & 0.02 \\ 
   \midrule 
\multicolumn{13}{c}{\textcolor{red}{\textbf{Sample size = 36}}} \\ 
\midrule
\multirow{12}{*}{0.5} & \multirow{2}{*}{(10 ; 10)} & B1 & 0.90 & 0.83 & 0.86 & 0.84 & 0.83 & 0.98 & 0.97 & 0.97 & 0.97 & 0.97 \\ 
   &  & B2 & 0.90 & 0.82 & 0.84 & 0.84 & 0.81 & 0.98 & 0.96 & 0.97 & 0.97 & 0.96 \\ 
   \cmidrule{2-13}
 & \multirow{2}{*}{(30 ; 10)} & B1 & 0.23 & 0.16 & 0.14 & 0.17 & 0.15 & 0.32 & 0.24 & 0.21 & 0.24 & 0.23 \\ 
   &  & B2 & 0.80 & 0.62 & 0.73 & 0.58 & 0.66 & 0.93 & 0.88 & 0.89 & 0.86 & 0.89 \\ 
   \cmidrule{2-13}
 & \multirow{2}{*}{(30 ; 30)} & B1 & 0.15 & 0.07 & 0.09 & 0.07 & 0.08 & 0.20 & 0.12 & 0.13 & 0.12 & 0.14 \\ 
   &  & B2 & 0.13 & 0.07 & 0.08 & 0.06 & 0.07 & 0.20 & 0.11 & 0.13 & 0.10 & 0.12 \\ 
   \cmidrule{2-13}
 & \multirow{2}{*}{(50 ; 10)} & B1 & 0.01 & 0.00 & 0.01 & 0.00 & 0.00 & 0.01 & 0.00 & 0.01 & 0.01 & 0.00 \\ 
   &  & B2 & 0.77 & 0.63 & 0.70 & 0.50 & 0.66 & 0.94 & 0.88 & 0.91 & 0.82 & 0.88 \\ 
   \cmidrule{2-13}
 & \multirow{2}{*}{(50 ; 30)} & B1 & 0.00 & 0.00 & 0.00 & 0.00 & 0.00 & 0.00 & 0.00 & 0.00 & 0.00 & 0.00 \\ 
   &  & B2 & 0.10 & 0.05 & 0.06 & 0.03 & 0.05 & 0.16 & 0.09 & 0.10 & 0.06 & 0.10 \\ 
   \cmidrule{2-13}
 & \multirow{2}{*}{(50 ; 50)} & B1 & 0.00 & 0.00 & 0.00 & 0.00 & 0.00 & 0.00 & 0.00 & 0.00 & 0.00 & 0.00 \\ 
   &  & B2 & 0.00 & 0.00 & 0.00 & 0.00 & 0.00 & 0.00 & 0.00 & 0.00 & 0.00 & 0.00 \\ 
   \midrule
\multirow{12}{*}{1.5} & \multirow{2}{*}{(10 ; 10)} & B1 & 0.93 & 0.75 & 0.85 & 0.79 & 0.77 & 0.99 & 0.95 & 0.98 & 0.97 & 0.96 \\ 
   &  & B2 & 0.92 & 0.72 & 0.83 & 0.78 & 0.74 & 0.98 & 0.93 & 0.97 & 0.95 & 0.94 \\ 
   \cmidrule{2-13}
 & \multirow{2}{*}{(30 ; 10)} & B1 & 0.34 & 0.14 & 0.14 & 0.17 & 0.13 & 0.47 & 0.21 & 0.22 & 0.25 & 0.21 \\ 
   &  & B2 & 0.86 & 0.56 & 0.72 & 0.48 & 0.63 & 0.96 & 0.83 & 0.89 & 0.81 & 0.87 \\ 
   \cmidrule{2-13}
 & \multirow{2}{*}{(30 ; 30)} & B1 & 0.27 & 0.06 & 0.09 & 0.06 & 0.08 & 0.34 & 0.12 & 0.13 & 0.11 & 0.14 \\ 
   &  & B2 & 0.24 & 0.06 & 0.08 & 0.06 & 0.09 & 0.36 & 0.10 & 0.12 & 0.10 & 0.12 \\ 
   \cmidrule{2-13}
 & \multirow{2}{*}{(50 ; 10)} & B1 & 0.04 & 0.01 & 0.00 & 0.01 & 0.01 & 0.05 & 0.01 & 0.01 & 0.01 & 0.01 \\ 
   &  & B2 & 0.84 & 0.55 & 0.70 & 0.29 & 0.61 & 0.96 & 0.83 & 0.91 & 0.70 & 0.86 \\ 
   \cmidrule{2-13}
 & \multirow{2}{*}{(50 ; 30)} & B1 & 0.02 & 0.00 & 0.00 & 0.00 & 0.00 & 0.03 & 0.00 & 0.00 & 0.00 & 0.00 \\ 
   &  & B2 & 0.19 & 0.05 & 0.06 & 0.01 & 0.06 & 0.27 & 0.08 & 0.10 & 0.04 & 0.11 \\ 
   \cmidrule{2-13}
 & \multirow{2}{*}{(50 ; 50)} & B1 & 0.01 & 0.00 & 0.00 & 0.00 & 0.00 & 0.02 & 0.00 & 0.00 & 0.00 & 0.00 \\ 
   &  & B2 & 0.01 & 0.00 & 0.00 & 0.00 & 0.00 & 0.02 & 0.00 & 0.00 & 0.00 & 0.00 \\ 
   \bottomrule
\end{tabularx}

\end{center}

\begin{center}
\small
\renewcommand{\arraystretch}{0.8}
\begin{tabularx}{\textwidth}{c c c *{10}{X}}
\caption{Probability of Early Termination (PET) under varying response rates, 
              two accrual rates, two interim analysis scenarios, 
              and sample sizes of 24 and 36 for three baskets.} \label{tab:pet2}\\
\toprule
\multirow{2}{*}{\makecell{\textbf{Accrual} \\ \textbf{Rate}}} &
\multirow{2}{*}{\makecell{\textbf{Response} \\ \textbf{Rate (B1, B2, B3)}}} &
\multirow{2}{*}{\textbf{Basket}} &
\multicolumn{5}{c}{\textbf{\# interim analyses = 1}} & \multicolumn{5}{c}{\textbf{\# interim analyses = 3}}
\\
\cmidrule(lr){4-8}\cmidrule(lr){9-13}
 &  &  & 
\textbf{NI} & \textbf{OD} & \textbf{CD} & \textbf{MI} & \textbf{MIC} & \textbf{NI} & \textbf{OD} & \textbf{CD} & \textbf{MI} & \textbf{MIC}
\endfirsthead

\toprule
\multirow{2}{*}{\makecell{\textbf{Accrual}\\\textbf{Rate}}} &
\multirow{2}{*}{\makecell{\textbf{Response}\\\textbf{Rate (B1, B2, B3)}}} &
\multirow{2}{*}{\textbf{Basket}} &
\multicolumn{5}{c}{\textbf{\# interim analyses = 1}} & \multicolumn{5}{c}{\textbf{\# interim analyses = 3}}
\\
\cmidrule(lr){4-8}\cmidrule(lr){9-13}
 &  &  & 
\textbf{NI} & \textbf{OD} & \textbf{CD} & \textbf{MI} & \textbf{MIC} & \textbf{NI} & \textbf{OD} & \textbf{CD} & \textbf{MI} & \textbf{MIC}
\\
\midrule
\endhead

\multicolumn{13}{r}{\small\textit{Continued on next page}}\\
\hline
\endfoot

\bottomrule
\endlastfoot
\bottomrule
\endlastfoot
\hline
\midrule 
\multicolumn{13}{c}{\textcolor{red}{\textbf{Sample size = 24}}} \\ 
\midrule
\multirow{30}{*}{0.5} & \multirow{3}{*}{(10 ; 10 ; 10)} & B1 & 0.86 & 0.76 & 0.81 & 0.77 & 0.74 & 0.96 & 0.92 & 0.94 & 0.93 & 0.91 \\ 
   &  & B2 & 0.86 & 0.75 & 0.79 & 0.76 & 0.74 & 0.98 & 0.94 & 0.94 & 0.95 & 0.92 \\ 
   &  & B3 & 0.86 & 0.75 & 0.80 & 0.76 & 0.74 & 0.96 & 0.91 & 0.94 & 0.92 & 0.90 \\ 
   \cmidrule{2-13}
 & \multirow{3}{*}{(30 ; 10 ; 10)} & B1 & 0.32 & 0.20 & 0.20 & 0.21 & 0.18 & 0.46 & 0.33 & 0.33 & 0.34 & 0.29 \\ 
   &  & B2 & 0.77 & 0.64 & 0.69 & 0.60 & 0.65 & 0.93 & 0.86 & 0.88 & 0.85 & 0.87 \\ 
   &  & B3 & 0.79 & 0.65 & 0.70 & 0.61 & 0.67 & 0.92 & 0.85 & 0.88 & 0.84 & 0.85 \\ 
   \cmidrule{2-13}
 & \multirow{3}{*}{(30 ; 30 ; 10)} & B1 & 0.23 & 0.14 & 0.13 & 0.12 & 0.13 & 0.34 & 0.23 & 0.21 & 0.21 & 0.21 \\ 
   &  & B2 & 0.23 & 0.13 & 0.12 & 0.12 & 0.12 & 0.34 & 0.20 & 0.19 & 0.18 & 0.19 \\ 
   &  & B3 & 0.68 & 0.49 & 0.59 & 0.40 & 0.52 & 0.85 & 0.75 & 0.77 & 0.70 & 0.77 \\ 
   \cmidrule{2-13}
 & \multirow{3}{*}{(30 ; 30 ; 30)} & B1 & 0.17 & 0.07 & 0.07 & 0.06 & 0.08 & 0.24 & 0.14 & 0.13 & 0.11 & 0.15 \\ 
   &  & B2 & 0.17 & 0.08 & 0.07 & 0.06 & 0.09 & 0.24 & 0.12 & 0.11 & 0.10 & 0.13 \\ 
   &  & B3 & 0.17 & 0.08 & 0.07 & 0.06 & 0.09 & 0.26 & 0.13 & 0.12 & 0.11 & 0.14 \\ 
   \cmidrule{2-13}
 & \multirow{3}{*}{(50 ; 10 ; 10)} & B1 & 0.04 & 0.02 & 0.01 & 0.02 & 0.01 & 0.06 & 0.03 & 0.01 & 0.03 & 0.02 \\ 
   &  & B2 & 0.75 & 0.61 & 0.66 & 0.48 & 0.62 & 0.93 & 0.85 & 0.88 & 0.80 & 0.85 \\ 
   &  & B3 & 0.76 & 0.63 & 0.68 & 0.48 & 0.64 & 0.91 & 0.85 & 0.88 & 0.81 & 0.85 \\ 
   \cmidrule{2-13}
 & \multirow{3}{*}{(50 ; 30 ; 10)} & B1 & 0.03 & 0.01 & 0.01 & 0.01 & 0.01 & 0.03 & 0.01 & 0.01 & 0.02 & 0.01 \\ 
   &  & B2 & 0.20 & 0.12 & 0.11 & 0.09 & 0.11 & 0.30 & 0.19 & 0.18 & 0.14 & 0.17 \\ 
   &  & B3 & 0.68 & 0.45 & 0.59 & 0.28 & 0.47 & 0.83 & 0.74 & 0.76 & 0.62 & 0.76 \\ 
   \cmidrule{2-13}
 & \multirow{3}{*}{(50 ; 30 ; 30)} & B1 & 0.02 & 0.01 & 0.00 & 0.01 & 0.01 & 0.02 & 0.01 & 0.00 & 0.01 & 0.01 \\ 
   &  & B2 & 0.14 & 0.08 & 0.07 & 0.05 & 0.08 & 0.19 & 0.11 & 0.10 & 0.07 & 0.12 \\ 
   &  & B3 & 0.15 & 0.07 & 0.07 & 0.04 & 0.07 & 0.22 & 0.11 & 0.11 & 0.07 & 0.12 \\ 
   \cmidrule{2-13}
 & \multirow{3}{*}{(50 ; 50 ; 10)} & B1 & 0.02 & 0.01 & 0.00 & 0.01 & 0.01 & 0.03 & 0.01 & 0.01 & 0.01 & 0.01 \\ 
   &  & B2 & 0.02 & 0.01 & 0.00 & 0.00 & 0.00 & 0.03 & 0.01 & 0.01 & 0.01 & 0.01 \\ 
   &  & B3 & 0.69 & 0.41 & 0.61 & 0.21 & 0.43 & 0.82 & 0.72 & 0.77 & 0.56 & 0.75 \\ 
   \cmidrule{2-13}
 & \multirow{3}{*}{(50 ; 50 ; 30)} & B1 & 0.01 & 0.00 & 0.00 & 0.00 & 0.00 & 0.02 & 0.00 & 0.00 & 0.00 & 0.00 \\ 
   &  & B2 & 0.01 & 0.01 & 0.00 & 0.00 & 0.01 & 0.02 & 0.01 & 0.00 & 0.00 & 0.01 \\ 
   &  & B3 & 0.14 & 0.05 & 0.08 & 0.02 & 0.06 & 0.19 & 0.09 & 0.09 & 0.04 & 0.10 \\ 
   \cmidrule{2-13}
 & \multirow{3}{*}{(50 ; 50 ; 50)} & B1 & 0.01 & 0.00 & 0.00 & 0.00 & 0.00 & 0.01 & 0.00 & 0.00 & 0.00 & 0.00 \\ 
   &  & B2 & 0.01 & 0.00 & 0.00 & 0.00 & 0.00 & 0.01 & 0.00 & 0.00 & 0.00 & 0.00 \\ 
   &  & B3 & 0.01 & 0.00 & 0.00 & 0.00 & 0.00 & 0.01 & 0.00 & 0.00 & 0.00 & 0.00 \\ 
   \midrule
\multirow{30}{*}{1.5} & \multirow{3}{*}{(10 ; 10 ; 10)} & B1 & 0.92 & 0.64 & 0.80 & 0.70 & 0.64 & 0.98 & 0.89 & 0.94 & 0.92 & 0.88 \\ 
   &  & B2 & 0.93 & 0.64 & 0.79 & 0.69 & 0.66 & 0.99 & 0.89 & 0.94 & 0.92 & 0.88 \\ 
   &  & B3 & 0.91 & 0.62 & 0.79 & 0.68 & 0.63 & 0.98 & 0.88 & 0.94 & 0.91 & 0.87 \\ 
   \cmidrule{2-13}
 & \multirow{3}{*}{(30 ; 10 ; 10)} & B1 & 0.51 & 0.17 & 0.20 & 0.22 & 0.17 & 0.66 & 0.29 & 0.32 & 0.36 & 0.26 \\ 
   &  & B2 & 0.88 & 0.54 & 0.68 & 0.49 & 0.60 & 0.98 & 0.80 & 0.88 & 0.78 & 0.83 \\ 
   &  & B3 & 0.88 & 0.51 & 0.70 & 0.46 & 0.56 & 0.97 & 0.78 & 0.88 & 0.78 & 0.81 \\ 
   \cmidrule{2-13}
 & \multirow{3}{*}{(30 ; 30 ; 10)} & B1 & 0.44 & 0.12 & 0.12 & 0.10 & 0.14 & 0.57 & 0.21 & 0.20 & 0.20 & 0.21 \\ 
   &  & B2 & 0.42 & 0.13 & 0.11 & 0.13 & 0.14 & 0.57 & 0.19 & 0.20 & 0.18 & 0.20 \\ 
   &  & B3 & 0.81 & 0.42 & 0.57 & 0.25 & 0.51 & 0.94 & 0.68 & 0.78 & 0.58 & 0.75 \\ 
   \cmidrule{2-13}
 & \multirow{3}{*}{(30 ; 30 ; 30)} & B1 & 0.35 & 0.09 & 0.07 & 0.06 & 0.11 & 0.49 & 0.13 & 0.13 & 0.11 & 0.17 \\ 
   &  & B2 & 0.36 & 0.09 & 0.07 & 0.06 & 0.11 & 0.50 & 0.12 & 0.11 & 0.09 & 0.15 \\ 
   &  & B3 & 0.34 & 0.08 & 0.08 & 0.05 & 0.11 & 0.47 & 0.12 & 0.12 & 0.09 & 0.16 \\ 
   \cmidrule{2-13}
 & \multirow{3}{*}{(50 ; 10 ; 10)} & B1 & 0.16 & 0.02 & 0.01 & 0.03 & 0.03 & 0.21 & 0.03 & 0.02 & 0.04 & 0.03 \\ 
   &  & B2 & 0.84 & 0.51 & 0.67 & 0.30 & 0.57 & 0.97 & 0.78 & 0.89 & 0.66 & 0.80 \\ 
   &  & B3 & 0.84 & 0.50 & 0.67 & 0.29 & 0.54 & 0.96 & 0.78 & 0.89 & 0.66 & 0.79 \\ 
   \cmidrule{2-13}
 & \multirow{3}{*}{(50 ; 30 ; 10)} & B1 & 0.14 & 0.02 & 0.01 & 0.02 & 0.02 & 0.16 & 0.02 & 0.01 & 0.02 & 0.02 \\ 
   &  & B2 & 0.37 & 0.11 & 0.10 & 0.06 & 0.12 & 0.51 & 0.17 & 0.18 & 0.09 & 0.17 \\ 
   &  & B3 & 0.79 & 0.41 & 0.59 & 0.11 & 0.48 & 0.92 & 0.66 & 0.75 & 0.42 & 0.72 \\ 
   \cmidrule{2-13}
 & \multirow{3}{*}{(50 ; 30 ; 30)} & B1 & 0.12 & 0.01 & 0.00 & 0.01 & 0.01 & 0.14 & 0.01 & 0.00 & 0.01 & 0.02 \\ 
   &  & B2 & 0.31 & 0.08 & 0.07 & 0.02 & 0.10 & 0.43 & 0.11 & 0.10 & 0.04 & 0.13 \\ 
   &  & B3 & 0.28 & 0.07 & 0.08 & 0.02 & 0.10 & 0.40 & 0.10 & 0.11 & 0.04 & 0.14 \\ 
   \cmidrule{2-13}
 & \multirow{3}{*}{(50 ; 50 ; 10)} & B1 & 0.11 & 0.01 & 0.00 & 0.00 & 0.02 & 0.14 & 0.02 & 0.01 & 0.01 & 0.02 \\ 
   &  & B2 & 0.10 & 0.02 & 0.00 & 0.00 & 0.02 & 0.14 & 0.02 & 0.01 & 0.01 & 0.02 \\ 
   &  & B3 & 0.75 & 0.40 & 0.62 & 0.04 & 0.46 & 0.89 & 0.64 & 0.77 & 0.28 & 0.70 \\ 
   \cmidrule{2-13}
 & \multirow{3}{*}{(50 ; 50 ; 30)} & B1 & 0.09 & 0.01 & 0.00 & 0.00 & 0.01 & 0.11 & 0.01 & 0.00 & 0.00 & 0.01 \\ 
   &  & B2 & 0.08 & 0.01 & 0.00 & 0.00 & 0.01 & 0.10 & 0.01 & 0.00 & 0.00 & 0.02 \\ 
   &  & B3 & 0.26 & 0.06 & 0.07 & 0.01 & 0.08 & 0.35 & 0.09 & 0.10 & 0.02 & 0.12 \\ 
   \cmidrule{2-13}
 & \multirow{3}{*}{(50 ; 50 ; 50)} & B1 & 0.08 & 0.01 & 0.00 & 0.00 & 0.01 & 0.09 & 0.01 & 0.00 & 0.00 & 0.01 \\ 
   &  & B2 & 0.07 & 0.01 & 0.00 & 0.00 & 0.01 & 0.08 & 0.01 & 0.00 & 0.00 & 0.02 \\ 
   &  & B3 & 0.06 & 0.01 & 0.00 & 0.00 & 0.01 & 0.07 & 0.01 & 0.00 & 0.00 & 0.01 \\ 
   \midrule 
\multicolumn{13}{c}{\textcolor{red}{\textbf{Sample size = 36}}} \\ 
\midrule
\multirow{30}{*}{0.5} & \multirow{3}{*}{(10 ; 10 ; 10)} & B1 & 0.94 & 0.89 & 0.91 & 0.90 & 0.88 & 0.99 & 0.98 & 0.99 & 0.99 & 0.98 \\ 
   &  & B2 & 0.94 & 0.88 & 0.90 & 0.90 & 0.87 & 0.99 & 0.98 & 0.98 & 0.98 & 0.97 \\ 
   &  & B3 & 0.94 & 0.89 & 0.92 & 0.90 & 0.88 & 0.99 & 0.98 & 0.99 & 0.98 & 0.98 \\ 
   \cmidrule{2-13}
 & \multirow{3}{*}{(30 ; 10 ; 10)} & B1 & 0.33 & 0.23 & 0.22 & 0.24 & 0.20 & 0.44 & 0.32 & 0.31 & 0.34 & 0.29 \\ 
   &  & B2 & 0.88 & 0.79 & 0.83 & 0.78 & 0.79 & 0.97 & 0.94 & 0.96 & 0.94 & 0.94 \\ 
   &  & B3 & 0.90 & 0.80 & 0.86 & 0.79 & 0.81 & 0.98 & 0.96 & 0.97 & 0.95 & 0.96 \\ 
   \cmidrule{2-13}
 & \multirow{3}{*}{(30 ; 30 ; 10)} & B1 & 0.24 & 0.15 & 0.16 & 0.14 & 0.14 & 0.32 & 0.23 & 0.22 & 0.21 & 0.21 \\ 
   &  & B2 & 0.22 & 0.13 & 0.14 & 0.12 & 0.12 & 0.34 & 0.22 & 0.23 & 0.22 & 0.21 \\ 
   &  & B3 & 0.81 & 0.66 & 0.76 & 0.60 & 0.68 & 0.94 & 0.90 & 0.91 & 0.88 & 0.91 \\ 
   \cmidrule{2-13}
 & \multirow{3}{*}{(30 ; 30 ; 30)} & B1 & 0.16 & 0.08 & 0.09 & 0.07 & 0.08 & 0.22 & 0.14 & 0.14 & 0.12 & 0.15 \\ 
   &  & B2 & 0.14 & 0.07 & 0.09 & 0.06 & 0.07 & 0.21 & 0.12 & 0.13 & 0.11 & 0.13 \\ 
   &  & B3 & 0.15 & 0.07 & 0.08 & 0.06 & 0.08 & 0.23 & 0.12 & 0.13 & 0.12 & 0.14 \\ 
   \cmidrule{2-13}
 & \multirow{3}{*}{(50 ; 10 ; 10)} & B1 & 0.02 & 0.01 & 0.01 & 0.01 & 0.01 & 0.03 & 0.01 & 0.01 & 0.02 & 0.01 \\ 
   &  & B2 & 0.87 & 0.78 & 0.82 & 0.72 & 0.77 & 0.97 & 0.95 & 0.97 & 0.94 & 0.95 \\ 
   &  & B3 & 0.88 & 0.78 & 0.86 & 0.74 & 0.79 & 0.97 & 0.96 & 0.97 & 0.94 & 0.96 \\ 
   \cmidrule{2-13}
 & \multirow{3}{*}{(50 ; 30 ; 10)} & B1 & 0.01 & 0.00 & 0.00 & 0.00 & 0.00 & 0.02 & 0.01 & 0.01 & 0.01 & 0.01 \\ 
   &  & B2 & 0.20 & 0.12 & 0.13 & 0.07 & 0.11 & 0.30 & 0.20 & 0.21 & 0.16 & 0.19 \\ 
   &  & B3 & 0.81 & 0.63 & 0.76 & 0.51 & 0.65 & 0.93 & 0.89 & 0.91 & 0.85 & 0.90 \\ 
   \cmidrule{2-13}
 & \multirow{3}{*}{(50 ; 30 ; 30)} & B1 & 0.00 & 0.00 & 0.00 & 0.00 & 0.00 & 0.01 & 0.00 & 0.00 & 0.00 & 0.00 \\ 
   &  & B2 & 0.12 & 0.06 & 0.07 & 0.04 & 0.06 & 0.17 & 0.10 & 0.10 & 0.07 & 0.11 \\ 
   &  & B3 & 0.13 & 0.06 & 0.07 & 0.04 & 0.06 & 0.19 & 0.10 & 0.11 & 0.08 & 0.11 \\ 
   \cmidrule{2-13}
 & \multirow{3}{*}{(50 ; 50 ; 10)} & B1 & 0.01 & 0.00 & 0.00 & 0.00 & 0.00 & 0.01 & 0.00 & 0.00 & 0.00 & 0.00 \\ 
   &  & B2 & 0.01 & 0.00 & 0.00 & 0.00 & 0.00 & 0.01 & 0.00 & 0.00 & 0.00 & 0.00 \\ 
   &  & B3 & 0.80 & 0.63 & 0.76 & 0.43 & 0.65 & 0.93 & 0.89 & 0.92 & 0.83 & 0.89 \\ 
   \cmidrule{2-13}
 & \multirow{3}{*}{(50 ; 50 ; 30)} & B1 & 0.00 & 0.00 & 0.00 & 0.00 & 0.00 & 0.00 & 0.00 & 0.00 & 0.00 & 0.00 \\ 
   &  & B2 & 0.00 & 0.00 & 0.00 & 0.00 & 0.00 & 0.00 & 0.00 & 0.00 & 0.00 & 0.00 \\ 
   &  & B3 & 0.10 & 0.05 & 0.05 & 0.02 & 0.06 & 0.16 & 0.09 & 0.08 & 0.04 & 0.09 \\ 
   \cmidrule{2-13}
 & \multirow{3}{*}{(50 ; 50 ; 50)} & B1 & 0.00 & 0.00 & 0.00 & 0.00 & 0.00 & 0.00 & 0.00 & 0.00 & 0.00 & 0.00 \\ 
   &  & B2 & 0.00 & 0.00 & 0.00 & 0.00 & 0.00 & 0.00 & 0.00 & 0.00 & 0.00 & 0.00 \\ 
   &  & B3 & 0.00 & 0.00 & 0.00 & 0.00 & 0.00 & 0.00 & 0.00 & 0.00 & 0.00 & 0.00 \\ 
   \midrule
\multirow{30}{*}{1.5} & \multirow{3}{*}{(10 ; 10 ; 10)} & B1 & 0.96 & 0.83 & 0.91 & 0.87 & 0.82 & 1.00 & 0.98 & 0.99 & 0.98 & 0.97 \\ 
   &  & B2 & 0.96 & 0.81 & 0.90 & 0.86 & 0.80 & 0.99 & 0.97 & 0.99 & 0.98 & 0.96 \\ 
   &  & B3 & 0.96 & 0.84 & 0.91 & 0.88 & 0.83 & 0.99 & 0.97 & 0.99 & 0.98 & 0.97 \\ 
   \cmidrule{2-13}
 & \multirow{3}{*}{(30 ; 10 ; 10)} & B1 & 0.45 & 0.20 & 0.22 & 0.26 & 0.18 & 0.58 & 0.31 & 0.31 & 0.36 & 0.28 \\ 
   &  & B2 & 0.92 & 0.71 & 0.83 & 0.69 & 0.74 & 0.98 & 0.92 & 0.96 & 0.92 & 0.93 \\ 
   &  & B3 & 0.94 & 0.73 & 0.86 & 0.71 & 0.76 & 0.99 & 0.93 & 0.97 & 0.94 & 0.94 \\ 
   \cmidrule{2-13}
 & \multirow{3}{*}{(30 ; 30 ; 10)} & B1 & 0.38 & 0.14 & 0.15 & 0.14 & 0.14 & 0.48 & 0.21 & 0.23 & 0.23 & 0.21 \\ 
   &  & B2 & 0.36 & 0.13 & 0.14 & 0.13 & 0.13 & 0.51 & 0.20 & 0.22 & 0.22 & 0.19 \\ 
   &  & B3 & 0.89 & 0.59 & 0.76 & 0.45 & 0.68 & 0.97 & 0.85 & 0.91 & 0.81 & 0.90 \\ 
   \cmidrule{2-13}
 & \multirow{3}{*}{(30 ; 30 ; 30)} & B1 & 0.29 & 0.07 & 0.09 & 0.06 & 0.09 & 0.38 & 0.12 & 0.13 & 0.11 & 0.15 \\ 
   &  & B2 & 0.28 & 0.07 & 0.09 & 0.06 & 0.10 & 0.42 & 0.10 & 0.13 & 0.10 & 0.14 \\ 
   &  & B3 & 0.30 & 0.08 & 0.08 & 0.06 & 0.09 & 0.41 & 0.13 & 0.14 & 0.11 & 0.15 \\ 
   \cmidrule{2-13}
 & \multirow{3}{*}{(50 ; 10 ; 10)} & B1 & 0.08 & 0.01 & 0.01 & 0.01 & 0.01 & 0.09 & 0.02 & 0.01 & 0.03 & 0.01 \\ 
   &  & B2 & 0.90 & 0.69 & 0.82 & 0.55 & 0.71 & 0.98 & 0.93 & 0.96 & 0.89 & 0.92 \\ 
   &  & B3 & 0.93 & 0.72 & 0.85 & 0.58 & 0.73 & 0.98 & 0.94 & 0.96 & 0.90 & 0.94 \\ 
   \cmidrule{2-13}
 & \multirow{3}{*}{(50 ; 30 ; 10)} & B1 & 0.04 & 0.01 & 0.00 & 0.01 & 0.01 & 0.06 & 0.01 & 0.01 & 0.01 & 0.01 \\ 
   &  & B2 & 0.31 & 0.11 & 0.12 & 0.06 & 0.11 & 0.43 & 0.18 & 0.20 & 0.13 & 0.17 \\ 
   &  & B3 & 0.88 & 0.56 & 0.76 & 0.30 & 0.64 & 0.96 & 0.85 & 0.90 & 0.73 & 0.88 \\ 
   \cmidrule{2-13}
 & \multirow{3}{*}{(50 ; 30 ; 30)} & B1 & 0.03 & 0.00 & 0.00 & 0.00 & 0.00 & 0.04 & 0.00 & 0.00 & 0.00 & 0.00 \\ 
   &  & B2 & 0.23 & 0.05 & 0.07 & 0.02 & 0.07 & 0.33 & 0.09 & 0.10 & 0.05 & 0.11 \\ 
   &  & B3 & 0.25 & 0.06 & 0.07 & 0.03 & 0.07 & 0.34 & 0.10 & 0.10 & 0.06 & 0.12 \\ 
   \cmidrule{2-13}
 & \multirow{3}{*}{(50 ; 50 ; 10)} & B1 & 0.04 & 0.00 & 0.00 & 0.00 & 0.00 & 0.05 & 0.00 & 0.00 & 0.00 & 0.00 \\ 
   &  & B2 & 0.04 & 0.00 & 0.00 & 0.00 & 0.00 & 0.04 & 0.01 & 0.00 & 0.00 & 0.00 \\ 
   &  & B3 & 0.86 & 0.55 & 0.76 & 0.17 & 0.62 & 0.95 & 0.85 & 0.92 & 0.65 & 0.87 \\ 
   \cmidrule{2-13}
 & \multirow{3}{*}{(50 ; 50 ; 30)} & B1 & 0.02 & 0.00 & 0.00 & 0.00 & 0.00 & 0.02 & 0.00 & 0.00 & 0.00 & 0.00 \\ 
   &  & B2 & 0.02 & 0.00 & 0.00 & 0.00 & 0.00 & 0.03 & 0.00 & 0.00 & 0.00 & 0.00 \\ 
   &  & B3 & 0.20 & 0.05 & 0.05 & 0.01 & 0.06 & 0.27 & 0.08 & 0.08 & 0.02 & 0.10 \\ 
   \cmidrule{2-13}
 & \multirow{3}{*}{(50 ; 50 ; 50)} & B1 & 0.01 & 0.00 & 0.00 & 0.00 & 0.00 & 0.01 & 0.00 & 0.00 & 0.00 & 0.00 \\ 
   &  & B2 & 0.01 & 0.00 & 0.00 & 0.00 & 0.00 & 0.02 & 0.00 & 0.00 & 0.00 & 0.00 \\ 
   &  & B3 & 0.01 & 0.00 & 0.00 & 0.00 & 0.00 & 0.02 & 0.00 & 0.00 & 0.00 & 0.00 \\ 
   \bottomrule
\end{tabularx}
\end{center}

\begin{center}
\small
\renewcommand{\arraystretch}{0.8}
\begin{tabularx}{\textwidth}{c c c *{10}{X}}
\caption{Average sample size 
              under varying response rates, 
              two accrual rates, two interim analysis scenarios, 
              and sample sizes of 24 and 36 for two baskets.}  \label{tab:ass_b2}\\
\toprule
\multirow{2}{*}{\makecell{\textbf{Accrual} \\ \textbf{Rate}}} &
\multirow{2}{*}{\makecell{\textbf{Response} \\ \textbf{Rate (B1, B2)}}} &
\multirow{2}{*}{\textbf{Basket}} &
\multicolumn{5}{c}{\textbf{\# interim analyses = 1}} & \multicolumn{5}{c}{\textbf{\# interim analyses = 3}}
\\
\cmidrule(lr){4-8}\cmidrule(lr){9-13}
 &  &  & 
\textbf{NI} & \textbf{OD} & \textbf{CD} & \textbf{MI} & \textbf{MIC} & \textbf{NI} & \textbf{OD} & \textbf{CD} & \textbf{MI} & \textbf{MIC}
\endfirsthead

\toprule
\multirow{2}{*}{\makecell{\textbf{Accrual}\\\textbf{Rate}}} &
\multirow{2}{*}{\makecell{\textbf{Response}\\\textbf{Rate (B1, B2)}}} &
\multirow{2}{*}{\textbf{Basket}} &
\multicolumn{5}{c}{\textbf{\# interim analyses = 1}} & \multicolumn{5}{c}{\textbf{\# interim analyses = 3}}
\\
\cmidrule(lr){4-8}\cmidrule(lr){9-13}
 &  &  & 
\textbf{NI} & \textbf{OD} & \textbf{CD} & \textbf{MI} & \textbf{MIC} & \textbf{NI} & \textbf{OD} & \textbf{CD} & \textbf{MI} & \textbf{MIC}
\\
\midrule
\endhead

\multicolumn{13}{r}{\small\textit{Continued on next page}}\\
\hline
\endfoot

\bottomrule
\endlastfoot
\bottomrule
\endlastfoot
\hline
\midrule 
\multicolumn{13}{c}{\textcolor{red}{\textbf{Sample size = 24}}} \\ 
 \midrule
\multirow{12}{*}{0.5} & \multirow{2}{*}{(10 ; 10)} & B1 & 14.49 & 15.85 & 15.21 & 15.62 & 15.70 & 13.54 & 14.56 & 14.07 & 14.34 & 14.51 \\ 
   &  & B2 & 14.36 & 15.90 & 15.30 & 15.80 & 15.78 & 13.33 & 14.45 & 14.06 & 14.28 & 14.45 \\ 
   \cmidrule{2-13}
 & \multirow{2}{*}{(30 ; 10)} & B1 & 21.18 & 22.22 & 22.32 & 22.20 & 22.32 & 20.28 & 21.74 & 21.61 & 21.73 & 21.92 \\ 
   &  & B2 & 15.41 & 18.39 & 16.38 & 18.88 & 17.85 & 14.54 & 16.52 & 15.53 & 16.96 & 16.09 \\ 
   \cmidrule{2-13}
 & \multirow{2}{*}{(30 ; 30)} & B1 & 21.83 & 23.14 & 22.96 & 23.24 & 23.09 & 21.49 & 22.76 & 22.72 & 22.88 & 22.66 \\ 
   &  & B2 & 22.09 & 23.11 & 22.92 & 23.18 & 23.00 & 21.76 & 22.89 & 22.74 & 22.99 & 22.80 \\ 
   \cmidrule{2-13}
 & \multirow{2}{*}{(50 ; 10)} & B1 & 23.68 & 23.84 & 23.94 & 23.86 & 23.88 & 23.60 & 23.83 & 23.92 & 23.84 & 23.88 \\ 
   &  & B2 & 15.53 & 18.44 & 16.60 & 20.66 & 18.08 & 14.64 & 16.72 & 15.60 & 18.32 & 16.50 \\ 
   \cmidrule{2-13}
 & \multirow{2}{*}{(50 ; 30)} & B1 & 23.78 & 23.93 & 23.94 & 23.92 & 23.93 & 23.76 & 23.92 & 23.94 & 23.90 & 23.91 \\ 
   &  & B2 & 22.33 & 23.11 & 23.14 & 23.60 & 23.06 & 22.00 & 22.94 & 22.94 & 23.49 & 22.84 \\ 
   \cmidrule{2-13}
 & \multirow{2}{*}{(50 ; 50)} & B1 & 23.83 & 23.96 & 23.96 & 23.99 & 23.96 & 23.82 & 23.96 & 23.96 & 23.98 & 23.96 \\ 
   &  & B2 & 23.82 & 23.95 & 23.96 & 23.99 & 23.95 & 23.82 & 23.94 & 23.96 & 23.99 & 23.94 \\ 
   \midrule
\multirow{12}{*}{1.5} & \multirow{2}{*}{(10 ; 10)} & B1 & 13.48 & 17.29 & 15.39 & 16.56 & 17.08 & 12.88 & 15.57 & 14.18 & 15.01 & 15.40 \\ 
   &  & B2 & 13.65 & 17.10 & 15.42 & 16.47 & 16.69 & 12.89 & 15.44 & 14.14 & 14.96 & 15.15 \\ 
   \cmidrule{2-13}
 & \multirow{2}{*}{(30 ; 10)} & B1 & 19.20 & 22.37 & 22.41 & 21.97 & 22.35 & 18.13 & 21.84 & 21.68 & 21.36 & 21.84 \\ 
   &  & B2 & 14.20 & 18.60 & 16.41 & 20.07 & 17.87 & 13.51 & 17.12 & 15.54 & 18.39 & 16.28 \\ 
   \cmidrule{2-13}
 & \multirow{2}{*}{(30 ; 30)} & B1 & 19.85 & 22.90 & 22.94 & 23.20 & 22.70 & 19.18 & 22.60 & 22.70 & 22.97 & 22.35 \\ 
   &  & B2 & 19.90 & 22.94 & 23.02 & 23.25 & 22.72 & 19.21 & 22.72 & 22.80 & 23.02 & 22.48 \\ 
   \cmidrule{2-13}
 & \multirow{2}{*}{(50 ; 10)} & B1 & 22.66 & 23.79 & 23.94 & 23.68 & 23.77 & 22.33 & 23.76 & 23.93 & 23.65 & 23.75 \\ 
   &  & B2 & 14.37 & 18.80 & 16.62 & 22.41 & 18.22 & 13.73 & 17.26 & 15.59 & 20.72 & 16.61 \\ 
   \cmidrule{2-13}
 & \multirow{2}{*}{(50 ; 30)} & B1 & 22.82 & 23.88 & 23.95 & 23.88 & 23.81 & 22.71 & 23.85 & 23.95 & 23.86 & 23.78 \\ 
   &  & B2 & 20.34 & 23.01 & 23.13 & 23.79 & 22.88 & 19.80 & 22.85 & 22.87 & 23.65 & 22.67 \\ 
   \cmidrule{2-13}
 & \multirow{2}{*}{(50 ; 50)} & B1 & 22.98 & 23.89 & 23.98 & 23.98 & 23.84 & 22.90 & 23.88 & 23.98 & 23.98 & 23.85 \\ 
   &  & B2 & 23.13 & 23.87 & 23.96 & 23.98 & 23.84 & 23.05 & 23.86 & 23.96 & 23.99 & 23.81 \\ 
   \midrule 
\multicolumn{13}{c}{\textcolor{red}{\textbf{Sample size = 36}}} \\ 
\midrule
\multirow{12}{*}{0.5} & \multirow{2}{*}{(10 ; 10)} & B1 & 19.82 & 21.08 & 20.59 & 20.88 & 21.04 & 18.93 & 19.66 & 19.34 & 19.48 & 19.64 \\ 
   &  & B2 & 19.88 & 21.32 & 20.96 & 20.96 & 21.50 & 19.02 & 19.93 & 19.67 & 19.69 & 19.95 \\ 
   \cmidrule{2-13}
 & \multirow{2}{*}{(30 ; 10)} & B1 & 31.85 & 33.11 & 33.40 & 32.95 & 33.29 & 30.90 & 32.31 & 32.70 & 32.22 & 32.52 \\ 
   &  & B2 & 21.61 & 24.81 & 22.94 & 25.57 & 24.06 & 20.32 & 22.36 & 21.41 & 22.81 & 21.80 \\ 
   \cmidrule{2-13}
 & \multirow{2}{*}{(30 ; 30)} & B1 & 33.34 & 34.76 & 34.33 & 34.71 & 34.55 & 32.81 & 34.24 & 33.95 & 34.22 & 34.01 \\ 
   &  & B2 & 33.68 & 34.72 & 34.49 & 34.85 & 34.70 & 33.07 & 34.36 & 34.11 & 34.52 & 34.22 \\ 
   \cmidrule{2-13}
 & \multirow{2}{*}{(50 ; 10)} & B1 & 35.80 & 35.95 & 35.89 & 35.93 & 35.96 & 35.78 & 35.94 & 35.89 & 35.91 & 35.96 \\ 
   &  & B2 & 22.08 & 24.69 & 23.45 & 26.99 & 24.20 & 20.61 & 22.29 & 21.61 & 23.99 & 21.94 \\ 
   \cmidrule{2-13}
 & \multirow{2}{*}{(50 ; 30)} & B1 & 35.95 & 36.00 & 36.00 & 36.00 & 36.00 & 35.94 & 36.00 & 36.00 & 36.00 & 36.00 \\ 
   &  & B2 & 34.26 & 35.08 & 34.96 & 35.44 & 35.03 & 33.76 & 34.69 & 34.65 & 35.19 & 34.59 \\ 
   \cmidrule{2-13}
 & \multirow{2}{*}{(50 ; 50)} & B1 & 35.98 & 36.00 & 36.00 & 36.00 & 36.00 & 35.98 & 36.00 & 36.00 & 36.00 & 36.00 \\ 
   &  & B2 & 36.00 & 36.00 & 36.00 & 36.00 & 36.00 & 36.00 & 36.00 & 36.00 & 36.00 & 36.00 \\ 
   \midrule
\multirow{12}{*}{1.5} & \multirow{2}{*}{(10 ; 10)} & B1 & 19.31 & 22.54 & 20.67 & 21.78 & 22.23 & 18.64 & 20.41 & 19.41 & 19.91 & 20.26 \\ 
   &  & B2 & 19.50 & 23.13 & 21.03 & 21.90 & 22.71 & 18.81 & 20.90 & 19.64 & 20.24 & 20.78 \\ 
   \cmidrule{2-13}
 & \multirow{2}{*}{(30 ; 10)} & B1 & 29.90 & 33.56 & 33.47 & 32.96 & 33.63 & 28.60 & 32.83 & 32.63 & 32.18 & 32.90 \\ 
   &  & B2 & 20.47 & 25.98 & 23.02 & 27.30 & 24.65 & 19.51 & 23.34 & 21.48 & 24.18 & 22.32 \\ 
   \cmidrule{2-13}
 & \multirow{2}{*}{(30 ; 30)} & B1 & 31.15 & 34.87 & 34.31 & 34.87 & 34.54 & 30.50 & 34.38 & 34.00 & 34.44 & 34.02 \\ 
   &  & B2 & 31.64 & 34.87 & 34.53 & 34.83 & 34.46 & 30.51 & 34.52 & 34.17 & 34.54 & 34.09 \\ 
   \cmidrule{2-13}
 & \multirow{2}{*}{(50 ; 10)} & B1 & 35.35 & 35.89 & 35.91 & 35.82 & 35.89 & 35.24 & 35.84 & 35.90 & 35.77 & 35.85 \\ 
   &  & B2 & 20.83 & 26.07 & 23.47 & 30.79 & 25.08 & 19.78 & 23.35 & 21.61 & 26.88 & 22.66 \\ 
   \cmidrule{2-13}
 & \multirow{2}{*}{(50 ; 30)} & B1 & 35.62 & 35.98 & 36.00 & 35.95 & 35.91 & 35.53 & 35.98 & 36.00 & 35.95 & 35.91 \\ 
   &  & B2 & 32.57 & 35.12 & 34.96 & 35.73 & 34.85 & 31.82 & 34.78 & 34.65 & 35.55 & 34.40 \\ 
   \cmidrule{2-13}
 & \multirow{2}{*}{(50 ; 50)} & B1 & 35.78 & 35.98 & 36.00 & 36.00 & 35.96 & 35.73 & 35.98 & 36.00 & 36.00 & 35.96 \\ 
   &  & B2 & 35.78 & 35.98 & 36.00 & 36.00 & 35.98 & 35.71 & 35.98 & 36.00 & 36.00 & 35.98 \\ 
   \bottomrule
\end{tabularx}

\label{tab:ass2}
\end{center}

\begin{center}
\small
\renewcommand{\arraystretch}{0.8}
\begin{tabularx}{\textwidth}{c c c *{10}{X}}
\caption{Average sample size 
              under varying response rates, 
              two accrual rates, two interim analysis scenarios, 
              and sample sizes of 24 and 36 for three baskets.}  \label{tab:ass_b3}\\
\toprule
\multirow{2}{*}{\makecell{\textbf{Accrual} \\ \textbf{Rate}}} &
\multirow{2}{*}{\makecell{\textbf{Response} \\ \textbf{Rate (B1, B2, B3)}}} &
\multirow{2}{*}{\textbf{Basket}} &
\multicolumn{5}{c}{\textbf{\# interim analyses = 1}} & \multicolumn{5}{c}{\textbf{\# interim analyses = 3}}
\\
\cmidrule(lr){4-8}\cmidrule(lr){9-13}
 &  &  & 
\textbf{NI} & \textbf{OD} & \textbf{CD} & \textbf{MI} & \textbf{MIC} & \textbf{NI} & \textbf{OD} & \textbf{CD} & \textbf{MI} & \textbf{MIC}
\endfirsthead

\toprule
\multirow{2}{*}{\makecell{\textbf{Accrual}\\\textbf{Rate}}} &
\multirow{2}{*}{\makecell{\textbf{Response}\\\textbf{Rate (B1, B2, B3)}}} &
\multirow{2}{*}{\textbf{Basket}} &
\multicolumn{5}{c}{\textbf{\# interim analyses = 1}} & \multicolumn{5}{c}{\textbf{\# interim analyses = 3}}
\\
\cmidrule(lr){4-8}\cmidrule(lr){9-13}
 &  &  & 
\textbf{NI} & \textbf{OD} & \textbf{CD} & \textbf{MI} & \textbf{MIC} & \textbf{NI} & \textbf{OD} & \textbf{CD} & \textbf{MI} & \textbf{MIC}
\\
\midrule
\endhead

\multicolumn{13}{r}{\small\textit{Continued on next page}}\\
\hline
\endfoot

\bottomrule
\endlastfoot
\bottomrule
\endlastfoot
\hline
\midrule 
\multicolumn{13}{c}{\textcolor{red}{\textbf{Sample size = 24}}} \\ 
 \midrule
 \midrule
\multirow{30}{*}{0.5} & \multirow{3}{*}{(10 ; 10 ; 10)} & B1 & 13.67 & 14.88 & 14.34 & 14.73 & 15.07 & 12.96 & 13.82 & 13.43 & 13.70 & 13.98 \\ 
   &  & B2 & 13.66 & 15.04 & 14.55 & 14.84 & 15.13 & 12.82 & 13.76 & 13.46 & 13.63 & 13.92 \\ 
   &  & B3 & 13.71 & 15.01 & 14.45 & 14.86 & 15.07 & 13.01 & 13.92 & 13.51 & 13.79 & 14.00 \\ 
   \cmidrule{2-13}
 & \multirow{3}{*}{(30 ; 10 ; 10)} & B1 & 20.20 & 21.54 & 21.55 & 21.48 & 21.80 & 19.21 & 20.73 & 20.75 & 20.61 & 21.12 \\ 
   &  & B2 & 14.77 & 16.32 & 15.74 & 16.75 & 16.20 & 13.69 & 14.88 & 14.51 & 15.19 & 14.83 \\ 
   &  & B3 & 14.55 & 16.21 & 15.61 & 16.74 & 16.00 & 13.67 & 14.87 & 14.38 & 15.21 & 14.78 \\ 
   \cmidrule{2-13}
 & \multirow{3}{*}{(30 ; 30 ; 10)} & B1 & 21.22 & 22.37 & 22.45 & 22.51 & 22.43 & 20.47 & 21.79 & 21.89 & 21.99 & 21.94 \\ 
   &  & B2 & 21.26 & 22.44 & 22.62 & 22.62 & 22.51 & 20.57 & 22.03 & 22.13 & 22.21 & 22.10 \\ 
   &  & B3 & 15.85 & 18.16 & 16.96 & 19.21 & 17.77 & 14.79 & 16.38 & 15.80 & 17.23 & 16.06 \\ 
   \cmidrule{2-13}
 & \multirow{3}{*}{(30 ; 30 ; 30)} & B1 & 22.01 & 23.10 & 23.10 & 23.26 & 22.99 & 21.49 & 22.71 & 22.76 & 22.95 & 22.58 \\ 
   &  & B2 & 22.01 & 23.03 & 23.21 & 23.22 & 22.97 & 21.55 & 22.78 & 22.95 & 23.00 & 22.67 \\ 
   &  & B3 & 21.99 & 23.05 & 23.10 & 23.25 & 22.93 & 21.46 & 22.73 & 22.79 & 22.95 & 22.65 \\ 
   \cmidrule{2-13}
 & \multirow{3}{*}{(50 ; 10 ; 10)} & B1 & 23.46 & 23.75 & 23.86 & 23.77 & 23.83 & 23.37 & 23.68 & 23.83 & 23.68 & 23.76 \\ 
   &  & B2 & 15.04 & 16.70 & 16.03 & 18.24 & 16.58 & 13.81 & 15.13 & 14.64 & 16.14 & 15.11 \\ 
   &  & B3 & 14.89 & 16.50 & 15.91 & 18.22 & 16.30 & 13.83 & 15.02 & 14.52 & 16.11 & 14.94 \\ 
   \cmidrule{2-13}
 & \multirow{3}{*}{(50 ; 30 ; 10)} & B1 & 23.68 & 23.84 & 23.90 & 23.84 & 23.88 & 23.64 & 23.83 & 23.90 & 23.82 & 23.85 \\ 
   &  & B2 & 21.59 & 22.56 & 22.73 & 22.97 & 22.68 & 20.96 & 22.11 & 22.21 & 22.68 & 22.29 \\ 
   &  & B3 & 15.87 & 18.63 & 16.87 & 20.69 & 18.39 & 14.98 & 16.69 & 15.92 & 18.44 & 16.43 \\ 
   \cmidrule{2-13}
 & \multirow{3}{*}{(50 ; 30 ; 30)} & B1 & 23.78 & 23.89 & 23.95 & 23.93 & 23.92 & 23.77 & 23.89 & 23.95 & 23.91 & 23.91 \\ 
   &  & B2 & 22.38 & 23.06 & 23.20 & 23.45 & 23.06 & 22.05 & 22.88 & 22.99 & 23.33 & 22.83 \\ 
   &  & B3 & 22.16 & 23.19 & 23.10 & 23.53 & 23.10 & 21.80 & 22.94 & 22.90 & 23.32 & 22.81 \\ 
   \cmidrule{2-13}
 & \multirow{3}{*}{(50 ; 50 ; 10)} & B1 & 23.72 & 23.88 & 23.94 & 23.93 & 23.90 & 23.64 & 23.86 & 23.93 & 23.92 & 23.90 \\ 
   &  & B2 & 23.72 & 23.88 & 23.94 & 23.95 & 23.94 & 23.65 & 23.86 & 23.91 & 23.93 & 23.90 \\ 
   &  & B3 & 15.77 & 19.11 & 16.65 & 21.48 & 18.89 & 14.95 & 17.07 & 15.70 & 19.39 & 16.79 \\ 
   \cmidrule{2-13}
 & \multirow{3}{*}{(50 ; 50 ; 30)} & B1 & 23.83 & 23.94 & 23.98 & 23.98 & 23.95 & 23.80 & 23.94 & 23.97 & 23.98 & 23.95 \\ 
   &  & B2 & 23.83 & 23.93 & 23.95 & 23.96 & 23.93 & 23.81 & 23.93 & 23.95 & 23.96 & 23.92 \\ 
   &  & B3 & 22.28 & 23.39 & 23.09 & 23.74 & 23.29 & 21.97 & 23.14 & 22.96 & 23.56 & 23.07 \\ 
   \cmidrule{2-13}
 & \multirow{3}{*}{(50 ; 50 ; 50)} & B1 & 23.83 & 23.96 & 23.96 & 23.99 & 23.96 & 23.83 & 23.96 & 23.96 & 23.99 & 23.96 \\ 
   &  & B2 & 23.83 & 23.96 & 23.96 & 24.00 & 23.96 & 23.83 & 23.96 & 23.96 & 24.00 & 23.96 \\ 
   &  & B3 & 23.88 & 23.99 & 23.95 & 23.99 & 23.98 & 23.88 & 23.98 & 23.95 & 23.99 & 23.96 \\ 
   \midrule
\multirow{30}{*}{1.5} & \multirow{3}{*}{(10 ; 10 ; 10)} & B1 & 12.94 & 16.39 & 14.43 & 15.61 & 16.30 & 12.55 & 14.75 & 13.54 & 14.12 & 14.73 \\ 
   &  & B2 & 12.91 & 16.38 & 14.55 & 15.72 & 16.15 & 12.44 & 14.74 & 13.50 & 14.22 & 14.71 \\ 
   &  & B3 & 13.18 & 16.59 & 14.53 & 15.81 & 16.49 & 12.63 & 14.94 & 13.55 & 14.29 & 14.87 \\ 
   \cmidrule{2-13}
 & \multirow{3}{*}{(30 ; 10 ; 10)} & B1 & 17.86 & 21.93 & 21.67 & 21.42 & 21.94 & 16.90 & 21.19 & 20.85 & 20.50 & 21.34 \\ 
   &  & B2 & 13.49 & 17.49 & 15.88 & 18.19 & 16.84 & 12.81 & 15.86 & 14.57 & 16.29 & 15.35 \\ 
   &  & B3 & 13.55 & 17.96 & 15.69 & 18.52 & 17.31 & 12.92 & 16.14 & 14.45 & 16.46 & 15.65 \\ 
   \cmidrule{2-13}
 & \multirow{3}{*}{(30 ; 30 ; 10)} & B1 & 18.79 & 22.54 & 22.51 & 22.74 & 22.38 & 17.90 & 22.02 & 21.97 & 22.17 & 21.89 \\ 
   &  & B2 & 18.98 & 22.48 & 22.67 & 22.50 & 22.36 & 18.01 & 22.09 & 22.14 & 22.16 & 21.95 \\ 
   &  & B3 & 14.35 & 18.98 & 17.18 & 21.06 & 17.92 & 13.52 & 17.30 & 15.94 & 19.01 & 16.33 \\ 
   \cmidrule{2-13}
 & \multirow{3}{*}{(30 ; 30 ; 30)} & B1 & 19.77 & 22.97 & 23.12 & 23.31 & 22.66 & 18.83 & 22.71 & 22.78 & 23.03 & 22.32 \\ 
   &  & B2 & 19.70 & 22.93 & 23.21 & 23.27 & 22.65 & 18.80 & 22.72 & 22.93 & 23.13 & 22.38 \\ 
   &  & B3 & 19.89 & 23.08 & 23.04 & 23.42 & 22.73 & 19.10 & 22.82 & 22.79 & 23.14 & 22.41 \\ 
   \cmidrule{2-13}
 & \multirow{3}{*}{(50 ; 10 ; 10)} & B1 & 22.07 & 23.70 & 23.87 & 23.63 & 23.69 & 21.68 & 23.67 & 23.83 & 23.55 & 23.65 \\ 
   &  & B2 & 13.95 & 17.86 & 16.02 & 20.44 & 17.24 & 13.09 & 16.14 & 14.62 & 18.23 & 15.72 \\ 
   &  & B3 & 13.97 & 18.04 & 15.99 & 20.58 & 17.59 & 13.13 & 16.26 & 14.51 & 18.23 & 15.98 \\ 
   \cmidrule{2-13}
 & \multirow{3}{*}{(50 ; 30 ; 10)} & B1 & 22.39 & 23.81 & 23.91 & 23.80 & 23.76 & 22.18 & 23.78 & 23.90 & 23.76 & 23.75 \\ 
   &  & B2 & 19.58 & 22.65 & 22.76 & 23.31 & 22.52 & 18.66 & 22.29 & 22.24 & 23.09 & 22.22 \\ 
   &  & B3 & 14.53 & 19.11 & 16.97 & 22.69 & 18.21 & 13.79 & 17.56 & 15.99 & 20.85 & 16.62 \\ 
   \cmidrule{2-13}
 & \multirow{3}{*}{(50 ; 30 ; 30)} & B1 & 22.61 & 23.93 & 23.95 & 23.92 & 23.84 & 22.43 & 23.89 & 23.95 & 23.90 & 23.82 \\ 
   &  & B2 & 20.25 & 23.00 & 23.19 & 23.70 & 22.77 & 19.52 & 22.82 & 23.02 & 23.59 & 22.58 \\ 
   &  & B3 & 20.60 & 23.12 & 23.02 & 23.71 & 22.83 & 19.91 & 22.94 & 22.84 & 23.60 & 22.57 \\ 
   \cmidrule{2-13}
 & \multirow{3}{*}{(50 ; 50 ; 10)} & B1 & 22.71 & 23.83 & 23.94 & 23.94 & 23.81 & 22.51 & 23.82 & 23.93 & 23.93 & 23.80 \\ 
   &  & B2 & 22.78 & 23.80 & 23.94 & 23.95 & 23.81 & 22.54 & 23.79 & 23.91 & 23.93 & 23.77 \\ 
   &  & B3 & 14.99 & 19.18 & 16.61 & 23.47 & 18.45 & 14.16 & 17.70 & 15.73 & 22.11 & 16.94 \\ 
   \cmidrule{2-13}
 & \multirow{3}{*}{(50 ; 50 ; 30)} & B1 & 22.90 & 23.92 & 23.98 & 23.98 & 23.86 & 22.77 & 23.90 & 23.98 & 23.97 & 23.85 \\ 
   &  & B2 & 23.03 & 23.88 & 23.95 & 23.98 & 23.82 & 22.87 & 23.87 & 23.95 & 23.96 & 23.80 \\ 
   &  & B3 & 20.91 & 23.22 & 23.11 & 23.92 & 23.02 & 20.36 & 23.05 & 22.96 & 23.83 & 22.77 \\ 
   \cmidrule{2-13}
 & \multirow{3}{*}{(50 ; 50 ; 50)} & B1 & 23.02 & 23.90 & 23.98 & 23.99 & 23.87 & 22.94 & 23.90 & 23.98 & 23.99 & 23.86 \\ 
   &  & B2 & 23.20 & 23.90 & 23.96 & 24.00 & 23.84 & 23.12 & 23.89 & 23.96 & 24.00 & 23.80 \\ 
   &  & B3 & 23.25 & 23.90 & 23.95 & 24.00 & 23.87 & 23.17 & 23.90 & 23.95 & 23.99 & 23.86 \\ 
   \midrule 
\multicolumn{13}{c}{\textcolor{red}{\textbf{Sample size = 36}}} \\ 
\midrule
\multirow{30}{*}{0.5} & \multirow{3}{*}{(10 ; 10 ; 10)} & B1 & 19.08 & 20.05 & 19.64 & 19.87 & 20.21 & 18.48 & 19.04 & 18.75 & 18.88 & 19.09 \\ 
   &  & B2 & 19.15 & 20.08 & 19.79 & 19.76 & 20.39 & 18.60 & 19.12 & 18.92 & 18.96 & 19.22 \\ 
   &  & B3 & 19.16 & 20.05 & 19.47 & 19.87 & 20.24 & 18.60 & 19.09 & 18.72 & 18.97 & 19.19 \\ 
   \cmidrule{2-13}
 & \multirow{3}{*}{(30 ; 10 ; 10)} & B1 & 30.10 & 31.88 & 32.06 & 31.65 & 32.42 & 29.03 & 30.98 & 31.11 & 30.69 & 31.55 \\ 
   &  & B2 & 20.14 & 21.77 & 21.14 & 21.97 & 21.72 & 19.23 & 20.22 & 19.84 & 20.36 & 20.26 \\ 
   &  & B3 & 19.89 & 21.58 & 20.44 & 21.77 & 21.34 & 19.00 & 19.96 & 19.36 & 20.12 & 19.90 \\ 
   \cmidrule{2-13}
 & \multirow{3}{*}{(30 ; 30 ; 10)} & B1 & 31.63 & 33.38 & 33.13 & 33.47 & 33.50 & 30.93 & 32.57 & 32.51 & 32.72 & 32.74 \\ 
   &  & B2 & 32.08 & 33.70 & 33.50 & 33.80 & 33.79 & 30.93 & 32.86 & 32.61 & 32.91 & 32.94 \\ 
   &  & B3 & 21.36 & 24.13 & 22.39 & 25.14 & 23.79 & 20.05 & 21.79 & 20.77 & 22.40 & 21.51 \\ 
   \cmidrule{2-13}
 & \multirow{3}{*}{(30 ; 30 ; 30)} & B1 & 33.09 & 34.62 & 34.33 & 34.76 & 34.58 & 32.49 & 34.05 & 33.93 & 34.24 & 33.92 \\ 
   &  & B2 & 33.48 & 34.79 & 34.40 & 34.85 & 34.69 & 32.77 & 34.33 & 33.99 & 34.41 & 34.18 \\ 
   &  & B3 & 33.37 & 34.69 & 34.56 & 34.83 & 34.58 & 32.54 & 34.18 & 34.03 & 34.31 & 33.93 \\ 
   \cmidrule{2-13}
 & \multirow{3}{*}{(50 ; 10 ; 10)} & B1 & 35.59 & 35.78 & 35.78 & 35.75 & 35.84 & 35.55 & 35.76 & 35.75 & 35.71 & 35.80 \\ 
   &  & B2 & 20.42 & 22.06 & 21.20 & 23.12 & 22.22 & 19.37 & 20.29 & 19.80 & 21.00 & 20.39 \\ 
   &  & B3 & 20.10 & 21.88 & 20.57 & 22.71 & 21.75 & 19.21 & 20.11 & 19.46 & 20.64 & 20.07 \\ 
   \cmidrule{2-13}
 & \multirow{3}{*}{(50 ; 30 ; 10)} & B1 & 35.77 & 35.91 & 35.93 & 35.91 & 35.91 & 35.71 & 35.88 & 35.90 & 35.89 & 35.90 \\ 
   &  & B2 & 32.42 & 33.84 & 33.73 & 34.69 & 34.06 & 31.45 & 33.10 & 32.95 & 33.87 & 33.30 \\ 
   &  & B3 & 21.50 & 24.69 & 22.26 & 26.81 & 24.24 & 20.21 & 22.05 & 20.83 & 23.44 & 21.76 \\ 
   \cmidrule{2-13}
 & \multirow{3}{*}{(50 ; 30 ; 30)} & B1 & 35.93 & 36.00 & 36.00 & 36.00 & 36.00 & 35.90 & 35.99 & 35.99 & 35.99 & 35.99 \\ 
   &  & B2 & 33.93 & 34.99 & 34.79 & 35.30 & 34.96 & 33.41 & 34.60 & 34.48 & 34.99 & 34.49 \\ 
   &  & B3 & 33.66 & 34.98 & 34.78 & 35.30 & 34.88 & 33.10 & 34.52 & 34.40 & 34.91 & 34.36 \\ 
   \cmidrule{2-13}
 & \multirow{3}{*}{(50 ; 50 ; 10)} & B1 & 35.80 & 35.95 & 35.93 & 36.00 & 35.98 & 35.77 & 35.92 & 35.93 & 36.00 & 35.98 \\ 
   &  & B2 & 35.89 & 35.96 & 35.95 & 35.96 & 35.96 & 35.86 & 35.94 & 35.93 & 35.94 & 35.94 \\ 
   &  & B3 & 21.52 & 24.61 & 22.33 & 28.26 & 24.33 & 20.30 & 21.98 & 20.87 & 24.36 & 21.78 \\ 
   \cmidrule{2-13}
 & \multirow{3}{*}{(50 ; 50 ; 30)} & B1 & 35.95 & 36.00 & 35.96 & 36.00 & 36.00 & 35.94 & 35.99 & 35.96 & 36.00 & 35.99 \\ 
   &  & B2 & 35.96 & 35.98 & 35.98 & 36.00 & 36.00 & 35.94 & 35.97 & 35.97 & 35.99 & 35.99 \\ 
   &  & B3 & 34.15 & 35.03 & 35.06 & 35.66 & 34.99 & 33.63 & 34.72 & 34.87 & 35.39 & 34.68 \\ 
   \cmidrule{2-13}
 & \multirow{3}{*}{(50 ; 50 ; 50)} & B1 & 35.98 & 36.00 & 36.00 & 36.00 & 36.00 & 35.98 & 36.00 & 36.00 & 36.00 & 36.00 \\ 
   &  & B2 & 36.00 & 36.00 & 36.00 & 36.00 & 36.00 & 35.99 & 36.00 & 36.00 & 36.00 & 36.00 \\ 
   &  & B3 & 35.96 & 35.98 & 35.98 & 35.98 & 35.98 & 35.96 & 35.98 & 35.98 & 35.98 & 35.98 \\ 
   \midrule
\multirow{30}{*}{1.5} & \multirow{3}{*}{(10 ; 10 ; 10)} & B1 & 18.72 & 21.14 & 19.65 & 20.31 & 21.30 & 18.33 & 19.51 & 18.79 & 19.14 & 19.66 \\ 
   &  & B2 & 18.80 & 21.48 & 19.86 & 20.52 & 21.56 & 18.45 & 19.85 & 18.97 & 19.33 & 19.93 \\ 
   &  & B3 & 18.84 & 20.88 & 19.63 & 20.25 & 21.15 & 18.42 & 19.50 & 18.82 & 19.18 & 19.75 \\ 
   \cmidrule{2-13}
 & \multirow{3}{*}{(30 ; 10 ; 10)} & B1 & 27.91 & 32.31 & 32.02 & 31.40 & 32.75 & 26.64 & 31.29 & 31.13 & 30.35 & 31.79 \\ 
   &  & B2 & 19.48 & 23.27 & 21.17 & 23.56 & 22.69 & 18.79 & 21.13 & 19.89 & 21.18 & 20.81 \\ 
   &  & B3 & 19.16 & 22.84 & 20.47 & 23.16 & 22.28 & 18.60 & 20.80 & 19.41 & 20.90 & 20.44 \\ 
   \cmidrule{2-13}
 & \multirow{3}{*}{(30 ; 30 ; 10)} & B1 & 29.20 & 33.48 & 33.29 & 33.57 & 33.48 & 28.20 & 32.76 & 32.60 & 32.70 & 32.84 \\ 
   &  & B2 & 29.50 & 33.72 & 33.58 & 33.66 & 33.70 & 28.05 & 32.97 & 32.72 & 32.84 & 33.06 \\ 
   &  & B3 & 20.08 & 25.44 & 22.36 & 27.90 & 23.88 & 19.24 & 22.80 & 20.80 & 24.30 & 21.61 \\ 
   \cmidrule{2-13}
 & \multirow{3}{*}{(30 ; 30 ; 30)} & B1 & 30.74 & 34.76 & 34.37 & 34.99 & 34.45 & 29.89 & 34.27 & 33.98 & 34.46 & 33.88 \\ 
   &  & B2 & 30.90 & 34.80 & 34.46 & 34.91 & 34.28 & 29.54 & 34.46 & 34.04 & 34.51 & 33.90 \\ 
   &  & B3 & 30.64 & 34.60 & 34.58 & 34.87 & 34.31 & 29.56 & 34.15 & 34.04 & 34.42 & 33.78 \\ 
   \cmidrule{2-13}
 & \multirow{3}{*}{(50 ; 10 ; 10)} & B1 & 34.55 & 35.80 & 35.78 & 35.73 & 35.80 & 34.41 & 35.74 & 35.77 & 35.60 & 35.78 \\ 
   &  & B2 & 19.86 & 23.52 & 21.24 & 26.05 & 23.31 & 19.02 & 21.21 & 19.86 & 22.78 & 21.14 \\ 
   &  & B3 & 19.32 & 23.14 & 20.70 & 25.64 & 22.96 & 18.79 & 20.86 & 19.57 & 22.46 & 20.77 \\ 
   \cmidrule{2-13}
 & \multirow{3}{*}{(50 ; 30 ; 10)} & B1 & 35.19 & 35.87 & 35.91 & 35.87 & 35.86 & 35.07 & 35.84 & 35.89 & 35.84 & 35.83 \\ 
   &  & B2 & 30.42 & 33.97 & 33.76 & 34.92 & 34.06 & 29.21 & 33.37 & 33.00 & 34.33 & 33.51 \\ 
   &  & B3 & 20.27 & 26.02 & 22.34 & 30.58 & 24.44 & 19.41 & 23.07 & 20.95 & 26.68 & 22.12 \\ 
   \cmidrule{2-13}
 & \multirow{3}{*}{(50 ; 30 ; 30)} & B1 & 35.50 & 35.96 & 36.00 & 35.96 & 35.93 & 35.39 & 35.96 & 35.98 & 35.96 & 35.92 \\ 
   &  & B2 & 31.80 & 35.14 & 34.80 & 35.66 & 34.69 & 30.80 & 34.76 & 34.47 & 35.37 & 34.31 \\ 
   &  & B3 & 31.52 & 34.92 & 34.74 & 35.53 & 34.69 & 30.66 & 34.57 & 34.45 & 35.24 & 34.22 \\ 
   \cmidrule{2-13}
 & \multirow{3}{*}{(50 ; 50 ; 10)} & B1 & 35.35 & 35.93 & 35.93 & 36.00 & 35.95 & 35.24 & 35.93 & 35.93 & 36.00 & 35.95 \\ 
   &  & B2 & 35.35 & 35.93 & 35.93 & 35.98 & 35.96 & 35.26 & 35.90 & 35.92 & 35.96 & 35.95 \\ 
   &  & B3 & 20.54 & 26.07 & 22.35 & 32.86 & 24.83 & 19.65 & 23.09 & 20.90 & 28.52 & 22.41 \\ 
   \cmidrule{2-13}
 & \multirow{3}{*}{(50 ; 50 ; 30)} & B1 & 35.69 & 35.98 & 35.96 & 36.00 & 35.98 & 35.61 & 35.98 & 35.96 & 36.00 & 35.98 \\ 
   &  & B2 & 35.60 & 35.98 & 35.98 & 35.98 & 35.96 & 35.54 & 35.97 & 35.98 & 35.98 & 35.95 \\ 
   &  & B3 & 32.41 & 35.08 & 35.07 & 35.87 & 34.96 & 31.75 & 34.79 & 34.87 & 35.75 & 34.57 \\ 
   \cmidrule{2-13}
 & \multirow{3}{*}{(50 ; 50 ; 50)} & B1 & 35.77 & 35.98 & 36.00 & 36.00 & 35.96 & 35.74 & 35.98 & 36.00 & 36.00 & 35.96 \\ 
   &  & B2 & 35.77 & 36.00 & 36.00 & 36.00 & 35.98 & 35.71 & 36.00 & 36.00 & 36.00 & 35.98 \\ 
   &  & B3 & 35.73 & 35.96 & 35.98 & 35.98 & 35.96 & 35.72 & 35.96 & 35.98 & 35.98 & 35.95 \\ 
   \bottomrule
\end{tabularx}

\label{tab:ass3}
\end{center}








\end{document}